\documentclass{article}

\usepackage{arxiv}

\usepackage[utf8]{inputenc} 
\usepackage[T1]{fontenc}    
\usepackage{hyperref}       
\usepackage{url}            
\usepackage{booktabs}       
\usepackage{amsfonts}       
\usepackage{nicefrac}       
\usepackage{microtype}      
\usepackage{lipsum}		
\usepackage{graphicx}
\usepackage{natbib}
\usepackage{doi}
\usepackage{subcaption}
\usepackage{amsmath}

\title{Multiscale Analysis of Woven Composites Using Hierarchical Physically Recurrent Neural Networks}

\date{} 					

\author{ \href{https://orcid.org/0000-0002-4930-000X}{\includegraphics[scale=0.06]{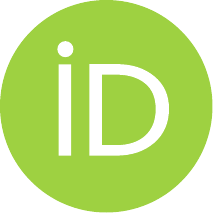}\hspace{1mm}Ehsan Ghane}\thanks{Corresponding author} \\ 
	Department of Physics\\
	University of Gothenburg\\
	Gothenburg, Sweden \\
	\texttt{ehsan.ghane@physics.gu.se} \\
	\And
	\href{https://orcid.org/0009-0005-5595-1203}{\includegraphics[scale=0.06]{orcid.pdf}\hspace{1mm}Marina A.~Maia} \\
    	Faculty of Civil Engineering and Geosciences \\
	Delft University of Technology\\
	Delft, The Netherlands \\
	\texttt{M.AlvesMaia@tudelft.nl} \\
	\AND
	\href{https://orcid.org/0000-0001-8410-3741}{\includegraphics[scale=0.06]{orcid.pdf}\hspace{1mm}Iuri B.~C.~M.~Rocha} \\
    	Faculty of Civil Engineering and Geosciences \\
	Delft University of Technology\\
	Delft, The Netherlands \\
	\texttt{I.Rocha@tudelft.nl} \\
	\And
	\href{https://orcid.org/0000-0001-7450-9086}{\includegraphics[scale=0.06]{orcid.pdf}\hspace{1mm}Martin Fagerström} \\
    	Department of Industrial and Materials Science \\
	Chalmers University of Technology\\
	Gothenburg, Sweden \\
	\texttt{martin.fagerstrom@chalmers.se} \\
	\And
	\href{https://orcid.org/0000-0000-0000-0000}{\includegraphics[scale=0.06]{orcid.pdf}\hspace{1mm}Mohsen Mirkhalaf} \\
    	Department of Physics\\
	University of Gothenburg\\
	Gothenburg, Sweden \\
	\texttt{mohsen.mirkhalaf@physics.gu.se} \\
}

\renewcommand{\shorttitle}{\textit{arXiv} HPRNN}

\hypersetup{
pdftitle={Multiscale Analysis of Woven Composites Using Hierarchical Physically Recurrent Neural Networks},
pdfsubject={q-bio.NC, q-bio.QM},
pdfauthor={David S.~Hippocampus, Elias D.~Striatum},
pdfkeywords={First keyword, Second keyword, More},
}

\begin{document}
\maketitle

\begin{abstract}
	Multiscale homogenization of woven composites requires detailed micromechanical evaluations, leading to high computational costs. Data-driven surrogate models based on neural networks address this challenge but often suffer from big data requirements, limited interpretability, and poor extrapolation capabilities.
    This study introduces a Hierarchical Physically Recurrent Neural Network (HPRNN) employing two levels of surrogate modeling. First, Physically Recurrent Neural Networks (PRNNs) are trained to capture the nonlinear elasto-plastic behavior of warp and weft yarns using micromechanical data. In a second scale transition, a physics-encoded meso-to-macroscale model integrates these yarn surrogates with the matrix constitutive model, embedding physical properties directly into the latent space. Adopting HPRNNs for both scale transitions can avoid nonphysical behavior often observed in predictions from pure data-driven recurrent neural networks and transformer networks. This results in better generalization under complex cyclic loading conditions. The framework offers a computationally efficient and explainable solution for multiscale modeling of woven composites.
\end{abstract}


\section{Introduction}
\label{Introduction}
Multiscale homogenization of woven composites presents challenges due to the complexity of their microstructural behavior and the transitions across micro-, meso-, and macroscales. At the microscale, the interactions between fibers and the matrix are critical, while the mesoscale considers bundles of fibers (yarns or tows) as a homogenized continuum, where the morphology of the woven structure is crucial. Finally, at the macroscale, the composite is studied at the lamina or laminate level as a homogeneous continuum \cite{liu2023extended}.

Numerous studies have proposed multiscale computational homogenization approaches, such as those reviewed in \cite{kaushik2022numerical}, to bridge microscale information to meso- and macroscale phenomena \cite{erol2017effects,calvo2023determination,patel2019compressive}. Techniques like FE\(^2\) rely on detailed micromechanical evaluations at the lower scales for every quadrature (Gauss) point and time step in the macroscale domain, which leads to prohibitive computational costs \cite{Geers2010, schroder2014numerical, Nguyen2012}. The nonlinear, path-dependent behavior of woven composites further complicates this challenge. Linking microscale information all the way to the macroscale would require two scale transitions (i.e., \(FE^3\) approach) involving extreme computational effort. These factors highlight the need for computationally efficient alternatives that can bypass the intensive costs associated with lower-scale homogenization while maintaining accuracy.

Recent studies propose approximate methods, also known as \textit{surrogate models}, for efficiently designing computer experiments. The objective is to create surrogate models that can predict the multiaxial stress state from a given history of multi-dimensional deformation, thereby avoiding the high computational cost of traditional simulations. These methods offer great potential for reducing the computational burden of multiscale analyses and can be broadly categorized into model hierarchy surrogates, data-driven surrogates and hybrid surrogates approaches.
\textit{Model hierarchy surrogates} maintain a physics foundation but simplify the system using techniques such as coarser discretization, relaxed convergence criteria, or reduced physical details. Examples include reduced-order modeling methods, which are tied to the underlying physics.
\textit{Data-driven surrogates} construct approximations by fitting high-fidelity microscale data using interpolation or regression techniques without embedding physical principles. Common examples include polynomial response surfaces, kriging (Gaussian process regression) \cite{mora2025operator}, neural networks (e.g., multi-layer perceptrons, also known as deep neural networks), radial basis functions \cite{yamanaka2023surrogate}, and splines \cite{eldred2004second, herrmann2024deep}.

Providing sufficient data for data-driven models in a supervised learning framework can be achieved via a well-planned design of experiment (DoE) or on-the-fly data generation \cite{rocha2021fly}. This approach enables the creation of surrogate models with precision in certain applications close to high-fidelity homogenizations at lower scales to address computational costs at upper scales simulations \cite{Wang2019, Liu2021, deng2024data, liu2024development}. For instance, \textit{feed-forward neural networks} have proven effective as surrogates, learning the complex relationships between microscale input parameters and homogenized mesoscale responses in woven composites \cite{ghane2023multiscale}.

Regarding nonlinear path-dependent phenomena, analogies can be drawn from the field of natural language processing in AI. Specifically, the similarities between hidden states in \textit{Recurrent Neural Networks (RNNs)} and the self-attention mechanism in \textit{transformers} with the internal variables governing thermodynamically irreversible processes have drawn significant attention \cite{buehler2024mechgpt, fuhg2024review, mirkhalaf2024micromechanics}. These similarities highlight the potential of neural network architectures to model complex material behaviors. Two main engineered formats of RNNs, i.e. GRU and LSTM and recently transformers \cite{zhongbo2024pre, hu2023deep} have been demonstrated to have an ability to capture nonlinear constitutive behaviors by mapping input features, such as deformation tensors, to relevant outputs, such as stress tensors, at both the micro- \cite{ghavamian2019accelerating} and mesoscale \cite{ghane2024recurrent}. More studies on constitutive modeling of composites using data-driven surrogates are reviewed in \cite{Liu2021, mirkhalaf2024micromechanics}.

One major limitation of conventional data-driven surrogate models, often called data-hungry models, is their heavy dependence on large and diverse datasets for accurate predictions \cite{fuhg2024review}. Acquiring or generating these datasets, especially for high-fidelity applications, requires significant computational effort, making this approach impractical for large-scale or complex simulations.
Multi-fidelity approaches aim to address this issue by combining data from multiple sources, such as low-cost simulations and \textit{few-shot} high-fidelity simulations, to reduce computational costs \cite{zeng2023improving, fuhg2023modular, jakeman2022adaptive, cheung2024multi}. Transfer learning further enhances the applicability of surrogate models in multiscale woven composite modeling \cite{ghane2024recurrent}. By leveraging knowledge from related datasets, transfer learning enables models to perform well on new datasets with limited high-fidelity data, creating robust surrogates that operate across different fidelity levels \cite{Zhuang2021, goswami2020transfer}.

Despite these advancements, traditional data-driven approaches, including physics-guided neural networks \cite{xu2024physics,wang2018multiscale,ghane2024recurrent}, often lack intrinsic knowledge of the underlying physics. This limitation makes them prone to errors during extrapolation \cite{Bessa2017} and limits their interpretability, leading to the so-called \textit{black-box problem} \cite{Willcox2021, fuhg2024review}. These challenges highlight the need for models integrating physics-based information to enhance reliability and clarity.
\noindent \textit{Hybrid surrogate approaches} combine data fitting with physical knowledge to address the limitations of exclusively data-driven neural network surrogate models. These methods integrate physics-based insights into neural networks and can be broadly classified into two categories:\\
\textit{Physics-informed Neural Networks (PiNNs).} These models incorporate knowledge of physical laws such as conservation principles, boundary conditions, and kinematic relations by adding new terms related to these constraints in the neural network's loss function. Initially developed for solving Partial Differential Equations (PDEs) \cite{raissi2024physics}, PiNNs have since been applied to constitutive material modeling \cite{haghighat2021physics, rezaei2023learning, henkes2022physics, idrissi2024multiscale, jiang2023physically} and finite element (FE) analysis \cite{rezaei2022mixed}. However, their effectiveness is highly sensitive to the relative weighting of loss terms, and they struggle with heterogeneous materials like composites \cite{dwivedi2021distributed}.
\\
\textit{Physics-encoded Neural Networks.} These models embed physics-based knowledge directly into the architecture, such as through custom neurons, layers, or constraints \cite{garanger2024symmetry}. For example, Rao et al. \cite{rao2021hard} used element-wise feature multiplication to create a recurrent \(\Pi\)-block, mimicking terms governing PDEs for an optimization problem. Spatial dependencies are modeled using convolutional layers or finite-difference filters, while temporal evolution is handled via a forward Euler scheme. Other physics-encoded networks, like neural operators \cite{chen2018neural, goswami2023physics, mora2025operator} and Fourier neural operators \cite{li2020fourier}, integrate physics through engineered units combined with multi-layer perceptrons. 
Furthermore, Mostajeran et al. \cite{mostajeran2024epi} introduced an elasto-plasticity-informed Chebyshev-based Kolmogorov-Arnold network \cite{liu2024kan} to achieve accurate and generalizable function approximation for nonlinear stress-strain relationships while using fewer parameters than standard multi-layer perceptrons. Compared to physics-informed neural networks that use multi-layer perceptrons, Kolmogorov-Arnold networks require significantly less data for calibration and discovery. 

While hybrid surrogate approaches enhance interpretability and generalization to unseen conditions, they are often constrained by large dimensionality and structured data requirements, making them less practical for complex multiscale heterogeneous materials \cite{faroughi2024physics}. Further research is needed to extend these methods to challenging multiscale composite modeling. Besides, traditional multiscale modeling approaches often struggle to fully account for the complex interactions between different scales in woven composites \cite{ivanov2020modeling}. Some models oversimplify the geometry of woven structures, neglecting critical details that influence the material's response under various loading conditions \cite{Couegnat2019, wei2024multiscale}. This gap highlights the need for a robust and efficient approach to address the hierarchical complexity of woven composites.

This study extends the Physically Recurrent Neural Network (PRNN) framework \cite{maia2023physically}, which belongs to the family of physics-encoded hybrid neural networks and was originally inspired by the works of \cite{liu2019deep,hernandez2017}. 
The PRNN embeds the constitutive relations of the microscale as a modified hidden layer in a multi-layer perceptron (a feed-forward neural network), creating physically recurrent blocks with internal variables tied to thermodynamic constraints. Physically recurrent blocks explicitly track internal variables, ensuring that path-dependent behaviors such as elasto-plasticity are accurately modeled rather than relying on implicit memory mechanisms as in conventional recurrent networks. PRNNs have demonstrated strong performance in prior works in terms of reducing the training data requirements, extrapolation to non-monotonic scenarios, and significant computational efficiency in FE\(^2\) applications on unidirectional composites \cite{maia2023physically}. Nevertheless, its application to complex hierarchical structures remains unexplored.

\begin{figure}
    \centering
    \includegraphics[width=0.6\linewidth]{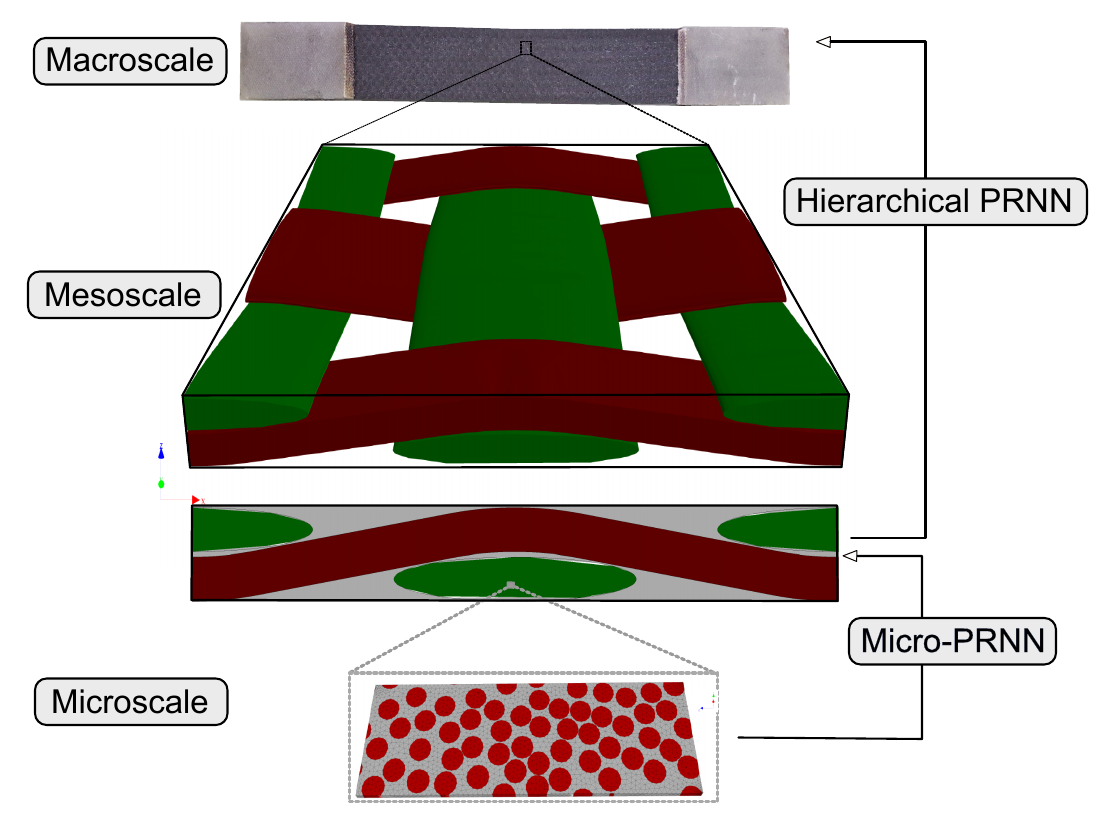}
    \caption{Hierarchical structure of woven composites and the transition across two scales with PRNN and the proposed HPRNN.}
    \label{fig:two-scale}
\end{figure}

A key innovation in this work is extending PRNNs to perform transitions in two scales (shown in Figure \ref{fig:two-scale}), effectively creating a surrogate model for \(FE^3\). While PRNNs were originally designed for a single scale transition, the current work treats a pretrained PRNN as a single material point at the mesoscale, allowing the construction of another PRNN for the second scale jump. Outside of directly solving \(FE^3\) or \textit{direct numerical simulations}, no existing surrogate method explicitly propagates microscopic constitutive behavior across two scales to the best of our knowledge.

Considering the material’s mesoscale structure, this work introduces a \textit{Hierarchical} PRNN (HPRNN) for the multiscale analysis of woven composites. 

This study assumes that non-linearity at the micro- and mesoscales arises from plasticity in the matrix, while the fibers exhibit isotropic elastic behavior. The modular nature of the proposed framework ensures that further developments, such as extending to more complex material behaviors or incorporating additional mechanisms, are easily accessible.

This paper is structured as follows. Section \ref{Background} discusses the challenges of computational homogenization for woven composites and reviews existing approaches. Section \ref{data_generation} describes the dataset used for training and evaluating the proposed models. Section \ref{PRNNarchitecture} provides an overview of PRNN design, as introduced in \cite{maia2023physically}. Section \ref{HPRNN_architecture} presents the HPRNN framework and its modular components. Section \ref{Results and discussion} evaluates HPRNN's performance against high-fidelity multiscale simulations and compares it with conventional history-dependent neural networks, including GRUs and Transformers, under elasto-plastic conditions. Finally, Section \ref{Conclusions} summarizes the key findings, highlights the advantages of HPRNN, and discusses its limitations and future research directions.

\section{Problem Overview}
\label{Background}

Woven fiber-reinforced polymer composites are specifically designed to conform more easily to complex curvatures than unidirectional laminates while  maintaining balanced and desirable mechanical properties \cite{ivanov2020modeling}. 
The plain woven fiber reinforcement pattern is tailored to provide balanced stiffness in both in-plane loading directions. However, this intricate morphology poses significant challenges when analyzing these materials. The mechanical response of woven composites varies depending on the loading direction. Under in-plane loading aligned with fiber bundles, the material exhibits high stiffness. In contrast, when the load is slightly misaligned with the fiber direction, the material demonstrates pronounced nonlinear behavior, even in systems comprising stiff carbon fibers and high-performance epoxy resins \cite{shokrieh2017general, taheri2022strength, ghanavaty2024progressive}.

Out-of-plane loading introduces additional complexities due to the interlaced fiber structure, especially in the absence of reinforcement fiber bundles in the out-of-plane direction. These conditions often lead to pronounced matrix-dominated behavior, including through-thickness deformation and potential delamination between yarn layers or fiber-matrix interfaces. The interfacial strength and the matrix’s capacity to absorb and redistribute stresses significantly influence the overall material response. Consequently, woven composites exhibit lower stiffness and strength in the out-of-plane direction than their in-plane properties.

Material nonlinearity arises from mechanisms such as (visco-)plasticity in the matrix phase or damage initiation and propagation in the matrix, yarns, or their interfaces. This study focuses solely on plasticity in the matrix phase, excluding rate-dependency and damage mechanisms. While yarns exhibit orthotropic behavior at the mesoscale, we assume the fibers are elastic and isotropic at the microscale. Although this assumption may lack physical accuracy, particularly for carbon fibers, it is reasonable for glass fibers and does not constrain the generality of our approach. We assume fibers to be homogeneous, disregarding their multilayer structure, and instead consider the fundamental fiber-matrix arrangement at the microscale. This simplification is justified, as microstructural observations indicate minimal fiber deformation and elastic behavior.

Understanding the material's loading history and constitutive behavior is essential to account for irreversible thermodynamic phenomena such as plasticity. This knowledge allows for the computation of accumulated internal variables, such as plastic strains. It also enables the prediction of plastic deformation using a yield surface and evolution rule, both of which depend on the chosen plasticity model.

A common approach in multiscale modeling is to formulate constitutive models directly at the mesoscale and assemble them using lamination theories. However, this often relies on phenomenological laws to capture complex material behavior, which can overlook key microscopic internal variables that ultimately govern the material's response. To address this limitation, we employ a computational homogenization framework at two levels that explicitly account for the microscale physics while bridging the gap to the mesoscale.

\subsection{Computational homogenization at two levels for woven composites}
\label{Multi-scale}

\begin{figure}
    \centering
    \includegraphics[width=0.8\linewidth]{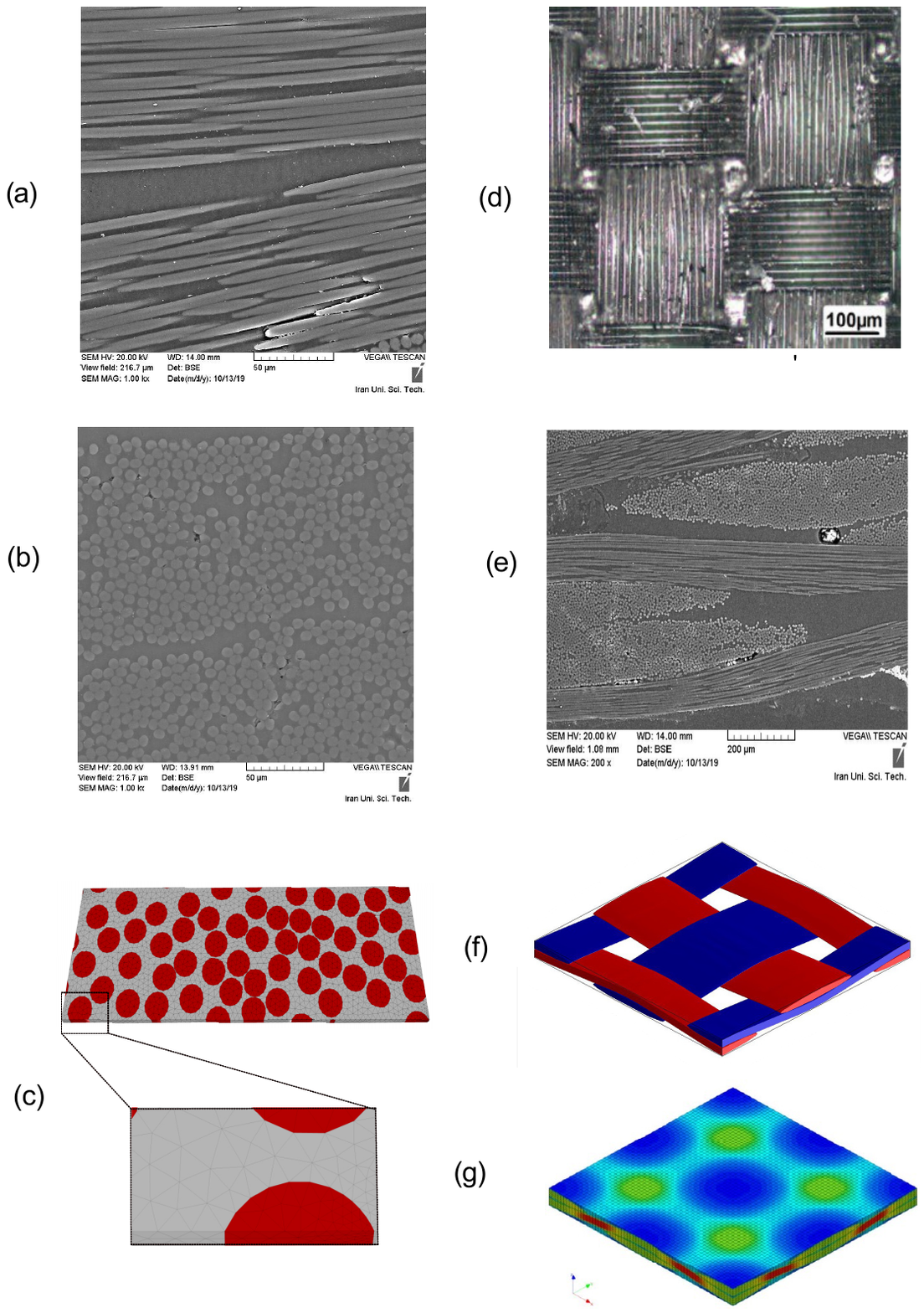}
    \caption{Illustration of woven composite structures and associated simulations. Microscale images captured by E. Ghane from \cite{ghane2020entropy} alongside equivalent representative volume elements (RVEs) modeled using Digimat-FE. The figure includes: (a,b) Scanning electron microscopy of fibers in two orthogonal directions (warp and weft), (c) microscale simulation of unidirectional composites, (d) top view microscopic image of a carbon fiber woven composite, (e) SEM image showing the cross-section of the woven composite, (f) mesoscale RVE, and (g) FEM simulation results with a voxel-based mesh.}
    \label{fig:multi_level}
\end{figure}

The two levels of the material under study, the microscale and mesoscale, are illustrated in the Figure \ref{fig:multi_level}.
Assuming periodicity \cite{Geers2010} at the subscale, the structural response is governed by effective properties derived from mesoscale analyses. Meanwhile, the microscale resolves the detailed behavior of the yarns' Representative Volume Element (RVE), capturing their intricate mechanical responses.

In first-order computational homogenization, scale separation is assumed, ensuring a uniform average deformation gradient across the RVE \cite{Geers2010}. The solution methodology typically involves iterative coupling between scales, with boundary conditions applied at subscale to maintain consistency. However, this nested nature introduces significant computational challenges, especially when solving the microscale RVE problem at every integration point within the mesoscale model. This complexity arises from the need to account for two scale transitions. First, from the microscale to the mesoscale, where fiber-matrix interactions define the yarns' effective behavior. Second, from the mesoscale to the macroscale, where the homogenized yarn response determines the overall composite properties while preserving microscale mechanisms like plasticity and damage evolution.

Machine learning techniques, like neural network-based models, can help solve these computational challenges in the micro-to-meso and meso-to-macro transition phases. Unlike conventional phenomenological approaches that impose constitutive assumptions at the mesoscale, data-driven methods can directly learn path-dependent responses from microscale simulations and incorporate them in meso-to-macro transition.

At the mesoscale, the constitutive responses of the yarns are not predefined. Instead, the mesoscopic behavior emerges naturally from solving the microscale boundary value problem. This eliminates the need for explicit assumptions regarding the constitutive behavior at the meso level, resulting in a more physically consistent representation of material behavior. Such a hierarchical approach is particularly well-suited for woven composites, where the interaction between matrix and yarns introduces substantial complexities across scales.

\subsection{Learning path-dependent material responses from data}
\label{path_dependent}
Capturing history-dependent material behavior requires models that can account for the sequential nature of stress-strain evolution. The goal is to develop surrogate models for the heterogeneous composite that predict the stress tensor based on a given history of multi-dimensional strain or deformation tensors, eliminating the need for computationally intensive simulations. Conventional feed-forward networks struggle with long-term dependencies \cite{li2023mechanics}, necessitating the use of sequential architectures such as Recurrent Neural Networks (RNNs) and transformers \cite{ghane2024recurrent, ghane2024datafusion, zhou2023one,lu2024fine,buehler2023melm}. However, these models heavily rely on extensive datasets to generalize well \cite{fuhg2024review}. This study extends the Physically Recurrent Neural Network (PRNN) framework \cite{maia2023physically} to overcome these limitations by incorporating physics-based constraints and explicit internal variables, ensuring robustness in extrapolation tasks.

\section{Data-generation}
\label{data_generation}
In this study, we generated two distinct datasets for model development. The first dataset is obtained from a high-fidelity microscale model that captures the elasto-plastic behavior of UD composites. This dataset with 500 samples is used for the micro- to mesoscale transition representing the warp and weft yarns. The second dataset with 400 samples is generated using Fast Fourier Transform (FFT) simulations for mesoscale homogenization of a woven composite RVE. Details on computational homogenization in both scales are described in Section \ref{mesocsaleModelingApproach}. Detailed material properties of the constituents are provided in Table \ref{tab:yield_parameters}, and the RVE geometrical descriptions are illustrated in Figure \ref{fig:computational_model}.

\subsection{Load generator algorithm}
\label{LoadGenerator}

A random-walk algorithm is utilized to generate six-dimensional input strain loading paths \cite{friemann2023micromechanics}. This algorithm updates the previous load step by adding two independent sources of white noise, both sampled from Gaussian noises with different scales. The first white noise is the primary driver on a larger scale (drift), while the second introduces small-scale variations (noise). Note that this algorithm differs from the load generator used in \cite{maia2023physically}, as it generates random steps independently without introducing any correlation between load steps.

\begin{figure}[ht]
    \centering
    \begin{subfigure}[b]{0.45\linewidth}
        \centering
        \includegraphics[width=0.8\linewidth]{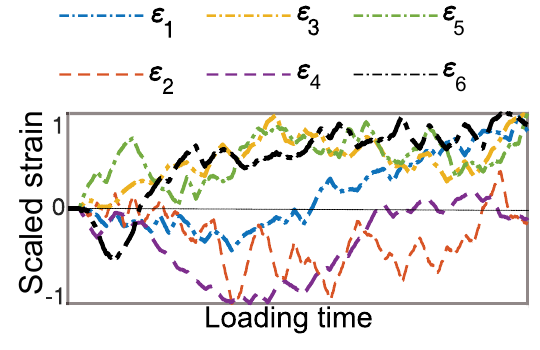}
        \caption{}
        \label{fig:random-stress12-a}
    \end{subfigure}
    \begin{subfigure}[b]{0.45\linewidth}
        \centering
        \includegraphics[width=0.8\linewidth]{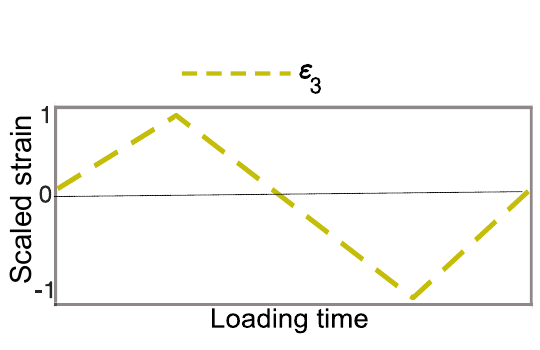}
        \caption{}
        \label{fig:random-stress12-b}
    \end{subfigure}
    \caption{Samples of scaled (normalized between 1 and -1) shear strain components $\varepsilon_{j}$ generated in (a) random loading used for training and validation and (b) cyclic loading to be used in extrapolation.}
    \label{fig:random-strain}
\end{figure}

The proposed data generation approach enables the generation of multi-axial stress-strain histories under random-walk loading conditions. To assess the trained networks' ability to extrapolate to sparse regions in the input space\footnote{The input space includes the six independent components of strain tensors. Sparsity refers to scenarios where specific strain components are zero, such as pure in-plane shear loading with only \(\varepsilon_{12}\) being nonzero.}, a second data generation strategy is also used \cite{ghane2024datafusion}. This second strategy focuses on cyclic in-plain shear loading with different peak strain values and number of cycles, where significant plasticity occurs in woven composites. Unlike random-walk loading, cyclic loading exhibits a strong correlation between consecutive strain values in a sequence. Consequently, cyclic loading is used as an extrapolation test case. Examples of shear strain components generated by the random-walk algorithm and cyclic load generation are shown in Figure \ref{fig:random-strain}.

\subsection{Material Models}
\label{material_models}

The framework relies on two primary constitutive models: an elasto-plastic model for the matrix and a linear elastic model for the fibers. These models are also used at the microscale during data generation and remain consistent across scales to ensure physical accuracy.

\subsubsection{Elasto-Plastic Matrix Model}\label{plastic}
\label{constitutiveMaterial}

The matrix material follows an elasto-plastic constitutive law based on $J_2$-plasticity, incorporating a linear-exponential hardening behavior as described in \cite{simo2006computational}. The yield function is defined as:

\begin{equation}
\Phi({\sigma}, \kappa) = \sigma_{\text{vM}} - \left(\sigma_{\text{y}} + H \bar{\varepsilon}^{\text{p}} + H_{\infty} \left(1 - e^{-m \bar{\varepsilon}^{\text{p}}}\right)\right) \leq 0,
\label{yield}
\end{equation}

where:
\begin{itemize}
\item[] $\sigma_{\text{vM}}$: von Mises stress
\item[] $\sigma_{\text{y}}$: initial yield stress
\item[] $H_{\infty}$: hardening modulus
\item[] $H$: linear hardening modulus
\item[] $m > 0$: hardening exponent
\item[] $\bar{\varepsilon}^{\text{p}} \geq 0$: accumulated plastic strain
\end{itemize}

The computational implementation follows the same return mapping algorithm used in the full-order micromodels, employing an iterative correction scheme to enforce consistency condition. The parameters, corresponding to a high-performance epoxy resin, are summarized in Table \ref{tab:yield_parameters}.

\subsubsection{Linear Elastic Fiber Model}\label{elastic}
The fiber reinforcements are assumed to behave elastically,
characterized by Young’s modulus $E$ and Poisson’s ratio $\nu$. The matrix and fibers' material parameters are presented in Table \ref{tab:yield_parameters}.

\begin{table}[h!]
\centering
\caption{Material parameters for the elasto-plastic matrix and elastic fiber models.}
\label{tab:yield_parameters}
\begin{tabular}{lcccccc}
\hline
\textbf{Material} & $\sigma_{\text{y}}$ [MPa] & $H$ [MPa] & $H_{\infty}$ [GPa] & $m$ [-] & $E$ [MPa] & $\nu$ [-] \\
\hline
Matrix & 18.00 & 10.00 & 65.00 & 180.00 & 3.7 & 0.35\\
Fiber & - & - & - & - & 275 & 0.22\\
\hline
\end{tabular}
\end{table}

\subsection{Computational homogenization of woven composites} 
\label{mesocsaleModelingApproach}

\begin{figure}
    \centering
    \includegraphics[width=0.6\linewidth]{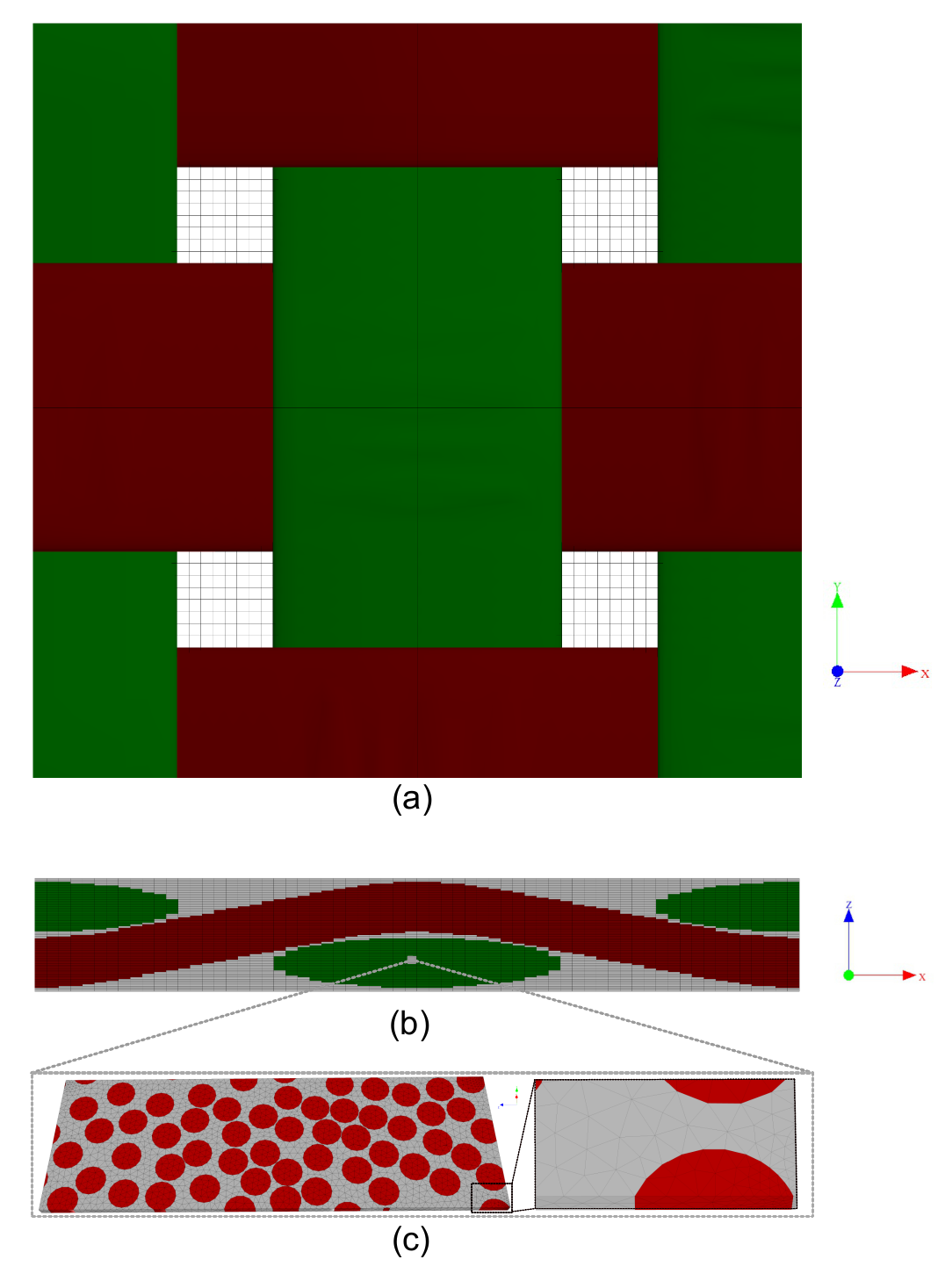}
    \caption{Details of the computational model at micro- and mesoscale: (a) Top view of the mesoscale model, (b) X-Z cross-section of the mesoscale model and the mesoscale mesh, and (c) the microscale warp yarn and the details on it mesh.}
    \label{fig:computational_model}
\end{figure}

Assuming no undulation through the yarns, microscale FE simulations are performed in Digimat-FE \cite{Digimat-FE} as shown in Figure \ref{fig:computational_model}(c).
The mesh configuration consists of a conforming tetrahedral mesh with an element size of \( 4.0 \, \mu m \) and a minimum element size of \( 0.2 \, \mu m \). Linear (first-order) elements are used, with internal coarsening and curvature control enabled to adapt the mesh to curved regions.

For mesoscale simulations, both FE and FFT methods are explored.
The RVE model used in FE and FFT processes represents a balanced weave with 15 yarns per centimeter in the warp and weft directions. Key geometric parameters include a yarn spacing ratio of 0.1, a crimp factor of 0.5, and an ellipsoidal yarn cross-section measuring 0.05~mm in height and 0.5~mm in width. 
The FE simulations are performed on a three-dimensional voxel-based mesh with dimensions 64~×~64~×~32 (A voxel represents a discrete computational element in each dimension of the simulation grid, with each voxel having dimensions of \(21\times21\times1.7\) \(\mu m\)), utilizing fully integrated 8-node elements. A convergence study verifies that further mesh refinement results in negligible changes to the stress-strain response, confirming the adequacy of this resolution. The FFT simulations employ a 64~×~64~×~64 voxel grid shown in Figure \ref{fig:computational_model}(a,b). Sensitivity analyses with higher grid resolutions show marginal improvements in accuracy. Thus, the resolution is found to be optimal by balancing computational efficiency and accuracy. A comparison between FE simulations with a voxel-based (non-conforming) mesh and FFT shows strong agreement in the results, with FFT-based reducing the computational cost of the stress-strain responses observed in voxel-based FE simulations \cite{ghane2024datafusion}. For instance, one FE simulation requires approximately 7040 seconds to converge during a cyclic loading, while FFT completes the task in 440 seconds. 

The RVE geometries (micro- and mesoscale) and the constitutive material models are kept fixed in both datasets. Simulations are conducted automatically with Julia \cite{bezanson2017julia} for writing input files and Digimat-FE \cite{Digimat-FE} for running on a system equipped with a 16-core processor and an \textit{NVIDIA\textregistered{} GeForce RTX\texttrademark{} 4090} GPU. The datasets vary based on the input loadings in terms of strain tensors, described in Section \ref{LoadGenerator}.
\section{Infusing physics-based recurrence into neural networks}
\label{PRNNarchitecture}

Developed by Maia et al. \cite{maia2023physically}, the Physically Recurrent Neural Network (PRNN) draws inspiration for its formulation and interpretation from computational homogenization principles. In the full-order solution, the mesoscale strain state is mapped onto local material points at the microscale, where classical constitutive models compute the stress-strain response, which is then homogenized back to the mesoscale. In the network, an analogous process takes place through an \textit{encoder-decoder} architecture in which the same constitutive models as in the full-order solution are embedded. The following section examines the role of each PRNN component in the first-scale transition, using terminology from computational homogenization. The second-scale transition is discussed in detail in Section \ref{HPRNN_architecture}.\\

\noindent \textbf{Localization}.
The localization or down-scaling step (analogous to an encoder) maps the mesoscopic strain inputs \( \boldsymbol{\varepsilon^\Omega} \) to a set of fictitious material points corresponding to microscale locations (global-to-local mapping). This step serves to approximate the microscopic boundary-value problem typically solved in finite element-based homogenization. The mapping is expressed as: 
\begin{equation}
\mathbf{a}_i = \phi(\mathbf{W}_i \mathbf{a}_{i-1} + \mathbf{b}_i), \quad i = 1, \dots, L
\end{equation}
where:
\begin{itemize}
    \item[] \( \mathbf{a}_0 = \boldsymbol{\varepsilon^\Omega} \) is the input mesoscopic strain at the current step,
    \item[] \( \mathbf{a}_i\) is the neurons states at layer \(i\) at the current step,
    \item[] \( \mathbf{W}_i \in \Re^{n_i\times n_{i-1}}\) and \( \mathbf{b}_i \) are the weight matrix and bias terms for layer \( i \) with \(n_i\) neurons,
    \item[] \( \phi(\cdot) \) is the (non)linear activation function, such as ReLU, sigmoid or tanh,
    \item[] \(L\) is the number of layers.
\end{itemize}

However, in this study, the encoder architecture consists of linear layers of activation functions, resulting in a direct relationship between mesoscopic and local strains. While the encoder-generated strain paths do not inherently contain history effects, the material response (stress) after the material layer exhibits path dependency, as in actual RVEs where history-dependent effects emerge due to nonlinear constitutive behavior. This path dependency is explicitly captured in the PRNN through history variables updated within the material layer.
It is important to note that the network does not store any displacement field data from the microscale model; the encoder instead learns mappings based exclusively on snapshots of macroscopic strains. This approximation enables the PRNN to reduce computational costs while retaining critical homogenized responses.

\noindent \textbf{Microscale constitutive response at material layer}.
The material layer consists of units representing fictitious material points designed to replicate the process of solving the microscale boundary-value problem in computational homogenization.

In each unit, a classical constitutive model \( \mathcal{D}^\omega \) maps the microscale strain \( \boldsymbol{\varepsilon^\omega} \) to the corresponding microscale stress \( \boldsymbol{\sigma^\omega} \) and internal variables \( \boldsymbol{\alpha}^\omega \) for that particular material point. This layer captures path-dependent behavior by updating the internal variables \( \boldsymbol{\alpha}^\omega_j \) at each time step, replacing (in combination with the encoder) the iterative solution of local equilibrium problems typically required in finite element homogenization.

For a given fictitious material point \( j \), the response is computed as:
\begin{equation}
\boldsymbol{\sigma}^\omega_j, \boldsymbol{\alpha}^\omega_{j,t} = \mathcal{D}^\omega \left( \boldsymbol{\varepsilon}^\omega_j, \boldsymbol{\alpha}^\omega_{j,t-1} \right),
\end{equation}
where:
\begin{itemize}
    \item[] \( \boldsymbol{\varepsilon}^\omega_j \) is the local strain at material point \( j \),
    \item[] \( \boldsymbol{\alpha}^\omega_{j,t-1} \) are the internal state variables, i.e., plastic strain, from time \( t-1 \),
    \item[] \( \mathcal{D}^\omega \) represents the constitutive model.
\end{itemize}

The formulation allows the network to adapt to a variety of nonlinear material behaviors, including elasto-plastic models with isotropic hardening or combinations of multiple constitutive models. For simplicity, this study models only the elasto-plastic matrix using a single material model \( \mathcal{D}^\omega \)
(Section \ref{plastic}), while the elastic fiber described in Section \ref{elastic} is not explicitly included in the network but is expected to be learned from the data. In other words, there is no explicitly defined linear elastic unit in the network architecture. Instead, the network infers and captures the fiber's elastic behavior based on the data it is trained on, as some training simulations will exhibit only elastic responses because they never reach the plastic regime, mimicking the elastic behavior of the composite. This constitutive model implemented in the micro-PRNNs is the same as that used for stress computations at the integration points of a full-order micro-model during the data generation phase. 

Instead of applying nonlinearity element-wise with scalar activation functions, neurons are grouped into sets, referred as fictitious material points, each corresponding to the number of strain components (i.e., size 6 for 3D analysis). While fewer in number than integration points in a finite element model, these fictitious material points serve as an efficient, interpretable representation of the material response, capturing key nonlinear behaviors. 
History dependency is inherently captured in the PRNN by storing and updating the stresses \( \boldsymbol{\sigma^\omega} \) and internal variables \( \boldsymbol{\alpha}^\omega \) associated with each fictitious material point.
The number of these fictitious material points is a hyperparameter and must be determined at each scale by a model selection procedure as explained in Section \ref{Results and discussion}.\\

\noindent \textbf{Homogenization.}
The decoder (analogous to the homogenization or up-scaling step) aggregates the stresses \( \boldsymbol{\sigma}^\omega \) computed at all fictitious material points to predict the mesoscopic stress \( \boldsymbol{\sigma^\Omega} \) (local-to-global integration.) This step mimics the homogenization operator in computational homogenization theory, where local fields are averaged to recover global responses. 
An important result of postulating the Hill-Mandel condition\footnote{The macroscopic volume average of the variation of work performed on the RVE is equal to the local variation of the work on the macroscale.} for an RVE with kinematic boundary conditions (fully prescribed or periodically tied\footnote{Composite material behavior under various loading conditions can be modeled using periodic boundary conditions, a realistic yet computationally efficient way of simulating materials with periodic microstructures. As compared to prescribed displacement or traction boundary conditions, these conditions provide a better estimate of the overall properties \cite{Geers2010,schroder2014numerical}.}) is that the macroscopic stress tensor \(\boldsymbol{\mathbf{\sigma}^\Omega}\) equals the volume average of the microscopic stress tensors \cite{Geers2010,schroder2014numerical}.
Similarly, in PRNN, the mesoscopic stress is computed as:

\begin{equation}
\widehat{\boldsymbol{\sigma}}^\Omega = \sum_{j=1}^m \mathbf{w}_j \boldsymbol{\sigma}^\omega_j,
\end{equation}
where \( \mathbf{w}_j \) are the neural network weights representing the contributions of the $m$ fictitious material points. To ensure non-negativity and physical consistency, the weights are defined using a \textit{SoftPlus} function:

\begin{equation}
\mathbf{w}_j = \log(1 + e^{\tilde{\mathbf{w}}_j}),
\end{equation}
where \( \tilde{\mathbf{w}}_j \) are unconstrained trainable parameters. The weights in the output layer represent the relative contribution of each fictitious material point to the macroscopic stress.
This step reconstructs the mesoscopic stress from the local responses, closing the computational homogenization loop.

\begin{figure}
    \centering
    \includegraphics[width=0.6\linewidth]{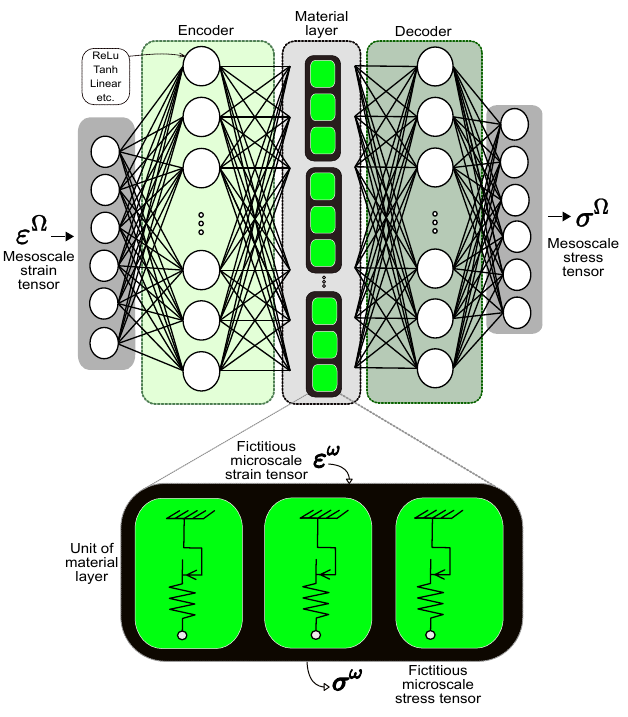}
    \caption{Architecture of PRNN used as the transition between micro- to mesoscales. It consists of a feed-forward encoder for strain decomposition, a material layer representing the constitutive material at the microscale yarn of the woven RVE, and a feed-forward decoder for homogenization. The schematics show three bulk points, but for the 3D tensors, six are considered as one unit.}
    \label{fig:prnn}
\end{figure}

\noindent \textbf{Training Process}.
The network is trained to minimize the standard mean square error loss function for each training path as follows:

\begin{equation}
\mathcal{L} = \frac{1}{N} \sum_{i=1}^N \frac{1}{2} \left\| \hat{\boldsymbol{\sigma}}^\Omega_i \left(\boldsymbol{\varepsilon}^\Omega_i \right) - \boldsymbol{\sigma}^\Omega_i \left(\boldsymbol{\varepsilon}^\Omega_i \right) \right\|^2,
\end{equation}
where:
\begin{itemize}{}
    \item[] \( {\sigma}^\Omega_i \) are the target (ground truth) mesoscopic stresses,
    \item[] \( {\hat\sigma}^\Omega_i \) are the predicted mesoscopic stresses,
    \item[] \( \boldsymbol{\varepsilon}^\Omega_i\) is the input strain at \(i^{th}\) snapshot,
    \item[] \( N \) is the number of snapshots (time steps).
\end{itemize}

Gradients are computed using backpropagation and optimized using a standard optimizer and automatic differentiation in PyTorch \cite{paszke2017automatic}.

\section{Hierarchical PRNN}\label{HPRNN_architecture}

Woven composites inherently exhibit a three-scale mechanical response as mentioned in Section \ref{Background}: the microscale (fiber-matrix interactions), the mesoscale (yarn structure), and the macroscale (composite response).

PRNNs direct application to woven composites faces a fundamental challenge. The constitutive response of yarns is not explicitly defined, and their complex interactions with the matrix make it challenging to construct a phenomenological material model.

To address this, we propose a hierarchical approach where PRNNs are first trained at the microscale to learn the nonlinear behavior of fiber-matrix interactions. It is anticipated that once trained, these microscale PRNNs can be frozen and employed as constitutive models at the mesoscale, enabling them to capture the inherent material behavior of yarns within the woven composite structure.

A key limitation of the frozen PRNN units is that they do not inherently account for yarns' orientations. Woven composites exhibit strong anisotropy due to their alternating warp and weft yarns. We introduce a rotation pre-processing and post-processing step around the PRNN modules to incorporate this directional dependence. This ensures the learned constitutive response is properly aligned with the local material orientation before homogenization.

\begin{figure}
    \centering
    \includegraphics[width=0.6\linewidth]{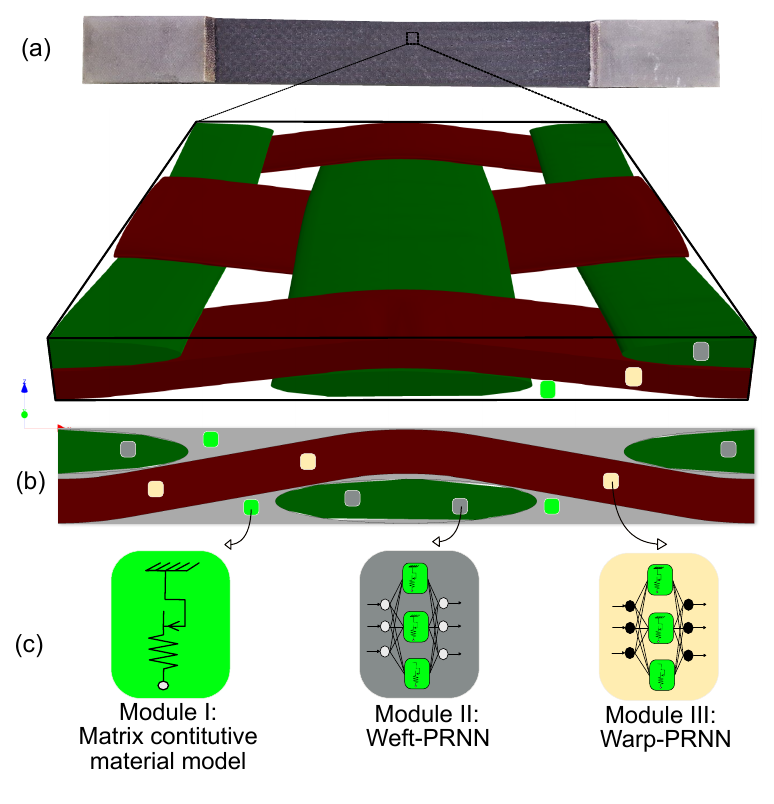}
    \caption{Multi-scale structure of a woven composite. (a) Photograph of a carbon fiber woven composite taken by E. Ghane, adapted from \cite{ghane2020entropy}, (b) Schematic representation of the mesoscale woven RVE, (c) The three fundamental components of a fictitious material point used in the HPRNN architecture: (I) matrix constitutive model, (II) weft-PRNN, and (III) warp-PRNN.}
    \label{fig:HPRNNMaterialLayer}
\end{figure}

With these modifications in place, we construct the Hierarchical PRNN (HPRNN), extending the standard PRNN framework to model woven composites. The architecture follows the principles of computational homogenization and consists of three key steps:
\begin{enumerate}
    \item \textit{Localization step}. Maps global strains to local strains within the matrix and yarns.
    \item \textit{Material layer}. Uses a combination of frozen PRNN modules for the yarns and a conventional constitutive model for the matrix.
    \item \textit{Homogenization step}. Calculates the mesoscopic stress field for the woven composite.
\end{enumerate}
The material layer, shown in Figure \ref{fig:HPRNNMaterialLayer}, consists of three modules: one representing the nonlinear response of the pure matrix phase (\textit{bulk material}), a \textit{warp-PRNN} trained at the microscale to model yarns in the longitudinal direction, and a \textit{weft-PRNN} trained at the microscale to model yarns in the transverse direction.

It is hypothesized that integrating these elements allows HPRNN to achieve an efficient yet physically consistent multiscale modeling strategy. The key hypothesis is that incorporating material-specific constitutive models into the material layer enhances the flexibility needed to capture the path-dependent behavior of woven composites while significantly reducing computational costs.

\subsection{Module I: Constitutive Material Model}
\label{constitutiveMaterialModule}

Module I represents the constitutive material model, which accounts for the behavior of the pure matrix phase and is directly used at the mesoscale. The matrix follows the elasto-plastic law described in Section~\ref{plastic}, while the fiber phase is assumed to be linearly elastic, as outlined in Section~\ref{elastic}. 

This module remains the most computationally intensive part of the framework since it requires solving an elasto-plastic return mapping problem at every time step. To ensure numerical stability, the same implementation used in the full-order micro-model is also adopted for this second scale transition. The return mapping algorithm iteratively enforces consistency condition, ensuring that the stress state remains on the yield surface.

\subsection{Module II: Warp PRNNs}
\label{warp}
To model the behavior of the warp phase, we treat the warp yarns as analogous to a Unidirectional (UD) composite ply, ignoring yarn undulations at local material points. 
The frozen PRNN, trained to capture the nonlinear behavior of a UD ply driven by matrix plasticity, is directly used as a surrogate for warp yarn behavior. This pre-trained network, obtained from micromechanical simulations of UD composites (Section \ref{data_generation}), effectively replicates the material's path-dependent response at the microscale. Its architecture follows the material layer design described in Section \ref{constitutiveMaterial}, with the number of fictitious material points and training dataset size validated in Section \ref{PRNN_performance}.

\subsection{Module III: Weft PRNNs}
\label{temporal correlation}
The orthogonality of warp and weft directions in a plain woven composite is handled through a standard tensor transformation. This allows the weft stress response to be directly derived from warp data, ensuring efficient characterization of the material's orthotropic behavior.

The transformation of a 3D second-order stress or strain tensor $\mathcal{T}$ is achieved using the standard tensor rotation formula \cite{kaw2005mechanics}:

\begin{equation}
    \mathcal{T}' = \mathcal{R} \, \mathcal{T} \, \mathcal{R}^T
    \label{eq:rotation_formula}
\end{equation}
where $\mathcal{T}$ is the original stress or strain tensor in the warp reference frame, $\mathcal{T}'$ is the transformed tensor in the weft frame, and $\mathcal{R}$ is the rotation matrix. For a 90$^\circ$ rotation around the z-axis (common in woven composites), the rotation tensor $\mathcal{R}$ is defined as \cite{daniel2006mechanics}:
\begin{equation}
    \mathcal{R} = 
    \begin{bmatrix}
    0 & -1 & 0 \\
    1 & 0 & 0 \\
    0 & 0 & 1 
    \end{bmatrix}
    \label{eq:rotation_matrix}
\end{equation}
This tensor rotates the x and y components while leaving the z-axis component unchanged. 

Equation~\eqref{eq:rotation_formula} enables estimating the weft stress-strain behavior from the warp data. 
A synthetic weft dataset is generated from the high-fidelity warp dataset by rotating the input strain and stress tensors. This approach allows for efficiently training a surrogate network for weft yarns, leveraging existing computational resources.

\subsection{Full Architecture Summary}\label{summary}
\begin{figure}
    \centering
    \includegraphics[width=0.6\linewidth]{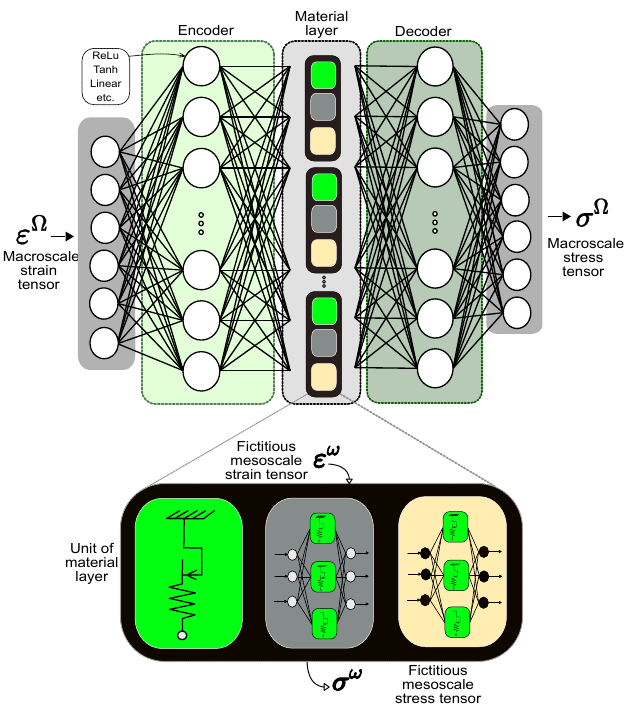}
    \caption{HPRNN architecture, featuring a standard feed-forward encoder for strain decomposition, a material layer representing the three primary phases of the material at the mesoscale of the woven RVE, and a feed-forward decoder for homogenization.}
    \label{fig:hprnnarchitecture}
\end{figure}

The HPRNN is designed to process six-dimensional strain input and predict six-dimensional stress outputs while incorporating material-specific microscale behavior through separate frozen PRNNs for warp and weft yarns. The architecture consists of three primary components according to Figure \ref{fig:hprnnarchitecture}: (1) an encoder that maps homogenized mesoscale strain inputs into latent representations, (2) a material layer that integrates the $J_2$ material model with warp and weft PRNN modules to capture the elasto-plastic behavior of yarns and interactions with the matrix, and (3) a decoder that maps the combined microscale fictitious stress responses back to the mesoscale output space.

Pre-trained PRNN modules are incorporated in evaluation mode (frozen as mentioned in Section \ref{HPRNN_architecture}), ensuring their weights remain fixed while optimizing the mesoscale network. The input data, comprising strain and stress tensors, is formatted into time series datasets for training and validation. The datasets are divided into sequences to ensure compatibility with the time-dependent architecture. The model processes all curves in the mini-batch for all time steps in a vectorized format, improving computational efficiency.

\noindent \textbf{Remark.} In contrast, the FFT simulations used for the meso-to-macroscale transition in Digimat-FE inherently rely on MFH for the microscale homogenization step. This is because the software must rapidly compute snapshots of effective material properties at each iteration. While MFH provides an efficient approximation, it does not fully capture the detailed stress and strain distributions within individual yarns, leading to potential inaccuracies when dealing with highly nonlinear material behaviors. However, the HPRNN framework differs in treating the micro-to-mesoscale transition at the yarn level and learns a direct transition from the microscale to the mesoscale.

This discrepancy introduces an interesting scenario (discussed further in Section \ref{PRNN_performance}): while we train HPRNN at the mesoscale using FFT simulations that depend on MFH for their lower-scale homogenization, the PRNN models within HPRNN perform a much more rigorous two-scale transition. In essence, HPRNN is tackling a significantly more complex problem by directly encoding the micro-to-mesoscale transition through finite-element-based data. However, due to computational infeasibility, we lack an FE² analysis to fully justify this approach at the macroscale.

Furthermore, for the meso-to-macroscale transition, HPRNN is trained on data generated from FFT simulations in Digimat-FE. Although these FFT simulations incorporate a simplification at the lower scale by relying on MFH, a plausible hypothesis is that the network effectively learns the macroscale behavior based on this data. Consequently, HPRNN predictions can still be meaningfully evaluated against FFT simulations at the macroscale, ensuring consistency with the multiscale computational framework.

\section{Results and discussion}
\label{Results and discussion}

In this section, we evaluate the performance of the HPRNN architectures proposed in Section \ref{HPRNN_architecture}. The model selection process is examined by analyzing the network's performance across training set sizes and material layer configurations. The evaluation specifically focuses on the ability of the HPRNN to accurately capture the homogenized response of woven composites under diverse loading conditions while incorporating microscale constitutive properties.

The PyTorch machine learning library is used to implement the HPRNN. Various configurations are trained on a 16-core computer locally. The training process for all the network configurations uses the ADAM optimizer with a learning rate of \(10\times 10^{-2}\). 
To mitigate overfitting, an early stopping criterion with a patience of 10 epochs is applied, meaning that training ends if the validation error does not improve for 10 consecutive epochs. 

Section \ref{PRNN_performance} presents the results related to the performance of the original PRNN architecture in capturing the homogenized yarn properties at the microscale. Section \ref{meso-PRNN} provides the results for the HPRNN architecture, which models the mesoscale behavior of woven composites. This is built upon the best-performing PRNNs selected in Section \ref{PRNN_performance} for modeling the microscale yarns' behavior. This section also compares the effectiveness of encoding physics-based recurrence into feed-forward networks against a conventional path-dependent GRU network, as explored in our previous work \cite{ghane2024datafusion} and a transformer-based network adopted from \cite{zhongbo2024pre}.
In Section \ref{extrapolation}, the ability of the HPRNN to extrapolate is tested using a cyclic loading scenario and compared with GRU performance. Finally, the discussion in Section \ref{discussion} highlights the advantages and limitations of the HPRNN architecture in the multiscale analysis of woven composites, providing insights into its potential applications and areas for improvement.

\subsection{Microscale yarn RVE performance}
\label{PRNN_performance}
We assess the performance of the original PRNN model in capturing the microscale elasto-plastic behavior of yarns using a newly generated random-walk dataset as described in Section \ref{mesocsaleModelingApproach}. The generated dataset features non-monotonic and non-proportional strain paths based on FE computational homogenization, providing a comprehensive training and test set to evaluate the micro-PRNNs' ability to generalize to complex loading scenarios. Each micro-to-meso PRNN is trained on 200 strain paths—a dataset size previously suggested by Maia et al. \cite{maia2023physically} as sufficient for capturing elasto-plasticity in unidirectional composites—then evaluated on 20 unseen samples.

\begin{figure}
    \centering
    \includegraphics[width=\linewidth]{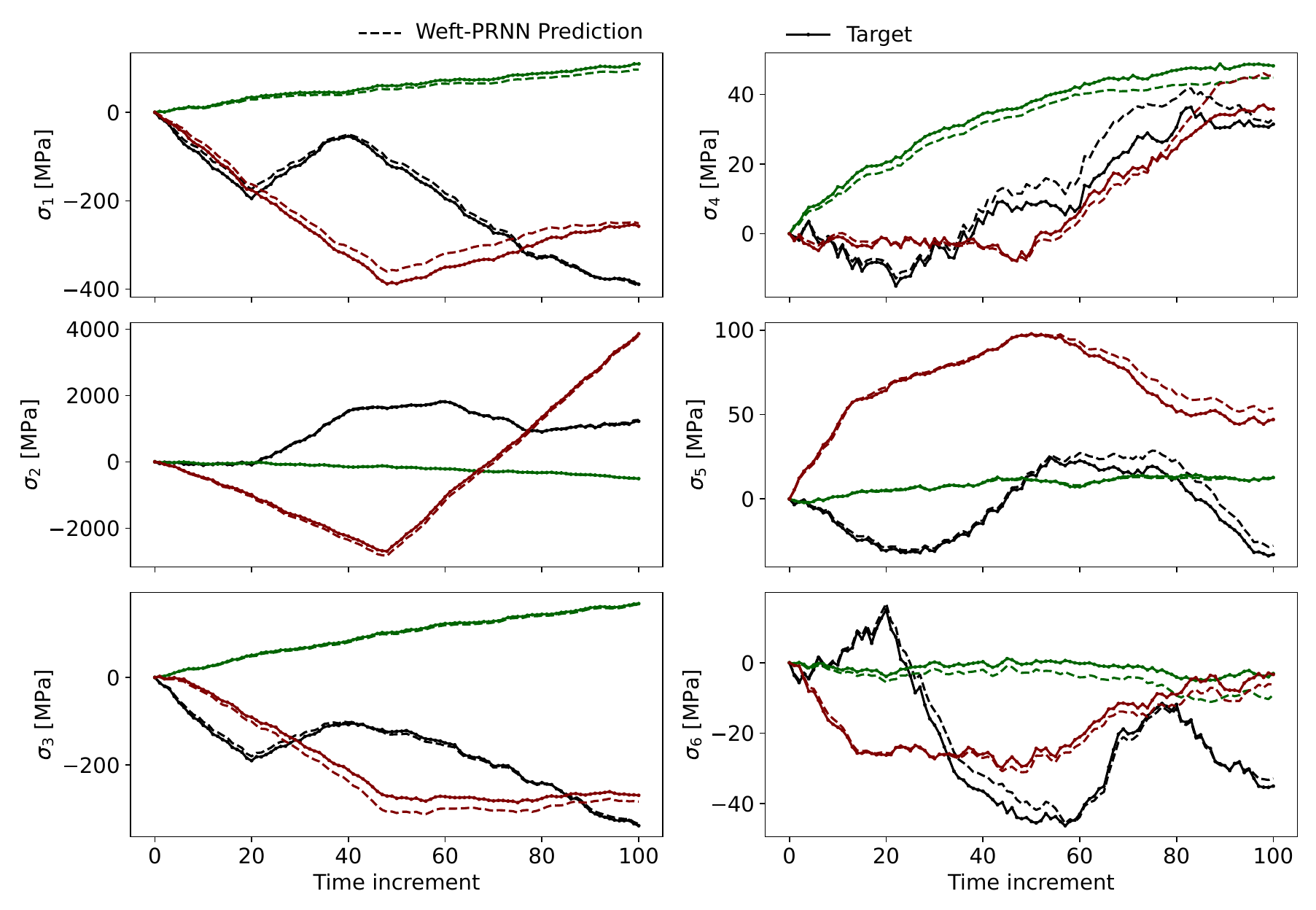}
    \caption{Prediction of stress components by the weft-PRNN with five bulk points for three randomly selected samples from the test set under random loading conditions.}
    \label{fig:enter-label}
\end{figure}

To ensure optimal performance, we conducted a sensitivity analysis across different configurations of bulk points and randomized initializations. However, since a similar analysis has already been performed by Maia et al. \cite{maia2023physically} on a different material system, we keep our discussion brief and do not include these results here. The effectiveness of this approach is further demonstrated in Figure \ref{fig:enter-label}, where PRNN successfully predicts the stress response\footnote{The symmetric stress tensor in Voigt notation is represented as a six-component column vector as follows: $\boldsymbol{\tilde{\sigma}} = (\sigma_{11}, \sigma_{22}, \sigma_{33}, \sigma_{12}, \sigma_{13}, \sigma_{23})^T \equiv (\sigma_{1}, \sigma_{2}, \sigma_{3}, \sigma_{4}, \sigma_{5}, \sigma_{6})^T$} under unseen random loading cases. 

\begin{figure}
        \centering
    \begin{subfigure}[b]{\linewidth}
        \centering
        \includegraphics[width=0.5\linewidth]{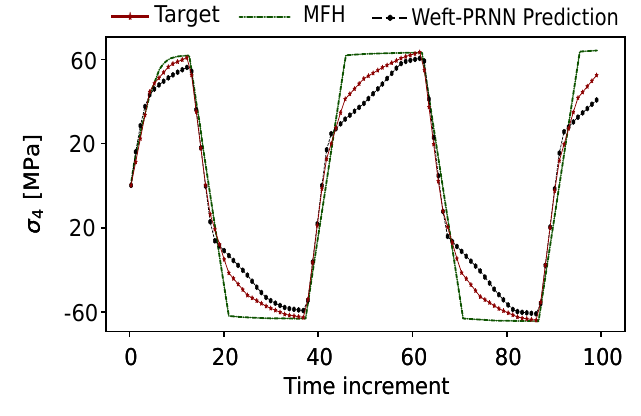}
        \label{fig:cyclicFEvsMGHvsPRNN}
        \caption{Cyclic in-plain shear loading.}
    \end{subfigure}
    \begin{subfigure}[b]{\linewidth}
        \centering
        \includegraphics[width=0.5\linewidth]{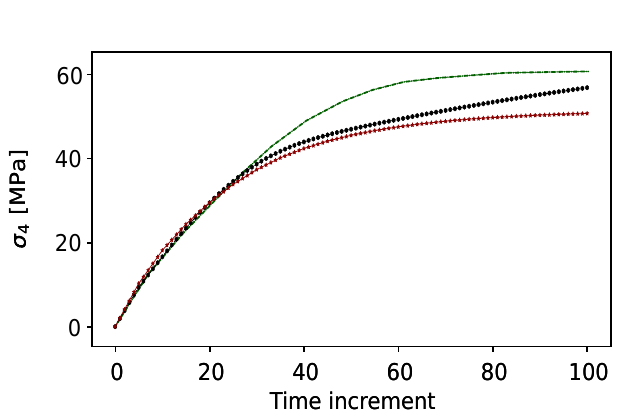}
        \label{fig:FEvsMGHvsPRNN}
        \caption{Uni-axial in-plain shear loading.}
    \end{subfigure}
    \caption{Comparison of micro-to-mesoscale transition for weft yarns under two standard loading scenarios not included in the training set.}
    \label{fig:mfh_fe_prnn_comparison}
\end{figure}

\noindent \textbf{Remark.} As outlined in Section \ref{summary}, Figure \ref{fig:mfh_fe_prnn_comparison} compares the micro-to-mesoscale transition for weft yarns under two standard loading scenarios that were not included in the training set. The results are presented for three different approaches:

\begin{itemize}
    \item \textit{Finite-Element-Based Full-order Homogenization.} Serves as the reference data for training warp-PRNN and weft-PRNN models, providing the \textit{target} stress responses.  
    \item \textit{PRNN Prediction.} Represents the output of the trained PRNN models, capturing the nonlinear and path-dependent behavior of yarns.  
    \item \textit{Mean-Field Homogenization (MFH) in Digimat-MF.} Used to homogenize the lower scale during FFT/FE simulations at the macroscale.  
\end{itemize}

The comparison shows that PRNN predictions align closely with FE results, accurately capturing the nonlinear and path-dependent behavior of yarns. PRNN successfully extrapolates from a random loading dataset to standard loading cases, demonstrating its robustness and generalization capability. However, the MFH approach shows larger deviations from FE results. This highlights the strength of HPRNN’s physics-encoded architecture, which effectively learns full-order homogenization without relying on simplified mean-field assumptions.

That said, this also means HPRNN is trained on potentially corrupted data, raising questions about the accuracy of results such as those in Section \ref{extrapolation}. Paradoxically, the surrogate model might be more accurate than the FE-based data used for training, which is an interesting aspect to consider in evaluating its predictive capabilities. While this limitation exists, we acknowledge it and proceed with caution in interpreting subsequent results.

\subsection{Mesoscale woven RVE Surrogate performance}
\label{meso-PRNN}

In the next phase, the mesoscale dataset is used to train and evaluate different neural network architectures for predicting the homogenized behavior of woven composites.

\begin{figure}
    \centering
    \includegraphics[width=0.6\linewidth]{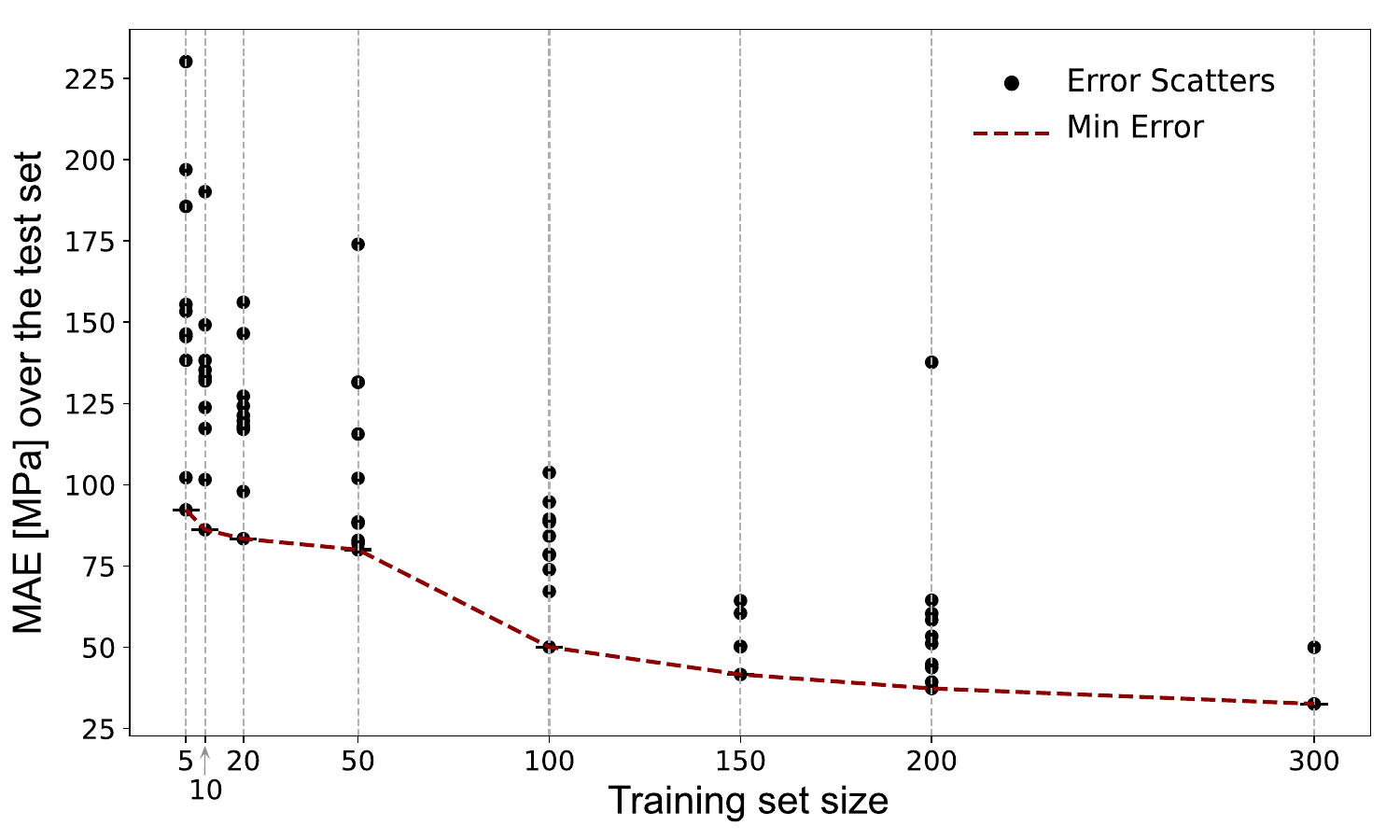}
    \caption{Comparison of von Mises stress mean absolute errors for the mesoscale network (HPRNN) configured with three latent space module sets, trained across varying training set sizes. Each training size involves training multiple networks (with the same configuration) with different random initializations, evaluated on a consistent set of 20 unseen simulations. The dashed line highlights the minimum errors achieved within each ensemble.}
    \label{fig:error_traiingSetSize}
\end{figure}

The optimum number of training set sizes is selected based on the minimum test error observed among the trained mesoscale networks according to Figure \ref{fig:error_traiingSetSize}. We have used the von Mises stress to incorporate all six stress components predicted by the network. This simplifies the assessment by consolidating multiple stress tensor components into a single scalar value with clear physical relevance. The Mean Absolute Error (MAE) measures the average magnitude of errors over the data sequence length ($N_T$) between the predicted and target values. MAE is calculated by taking the average of the absolute differences between each predicted and desired von Mises stress value, normalized by the number of tested samples from the unseen dataset ($M$) as:
\begin{equation}
\label{eq:MAE}
    \text{{MAE}} = \frac{1}{MN_T}\sum_{i=1}^{M}\sum_{t=1}^{N_T} \left|\hat\sigma_{{vM},i}^{(t)}-{\sigma}_{{vM},i}^{(t)}\right|,
\end{equation}

\noindent where $\hat{\sigma}^{(t)}_{vM}$ is the predicted von Mises stress at time step $t$ and $\sigma^{(t)}_{vM}$ is the desired von Mises stress at $t$. MAE indicates the average size of errors produced by the model. 

In this case, the networks with 200 and 300 training set sizes exhibited the lowest error values. While increasing the training set size generally improves model performance, it also significantly impacts computational cost. On a CPU with 32 processors, training mesoscale networks with 300 samples takes approximately 7 to 13 hours, while training with 200 samples takes approximately 4 hours. Given this trade-off between computational time and model accuracy, the candidate model is chosen by balancing performance gains against training costs. While the reduction in error achieved with 300 samples justifies the extra computational effort. Otherwise, the model trained with 200 samples is preferred for its efficiency.

\begin{figure}
    \centering
    \begin{subfigure}[b]{0.5\linewidth}
        \centering
        \includegraphics[width=\linewidth]{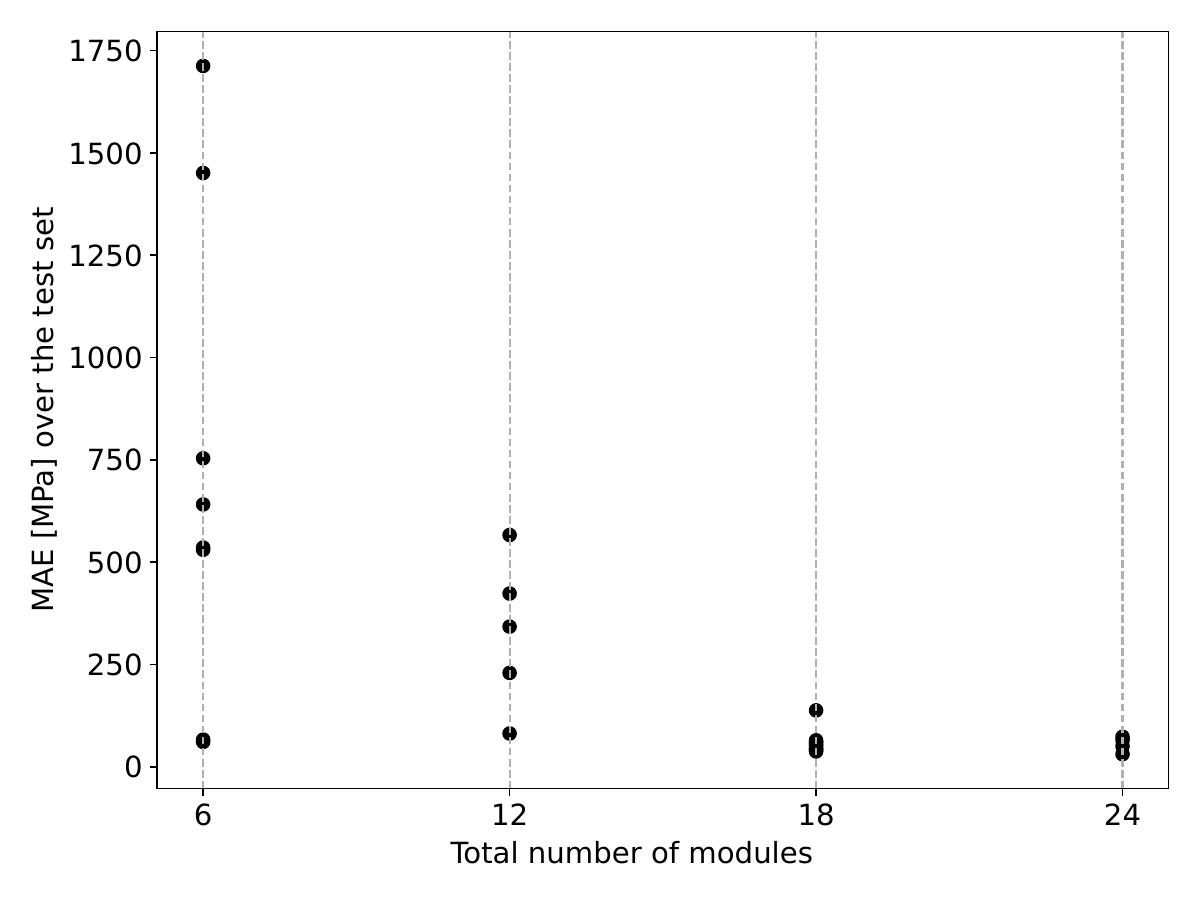}
        \caption{}
    \end{subfigure}
    \begin{subfigure}[b]{0.5\linewidth}
        \centering
        \includegraphics[width=\linewidth]{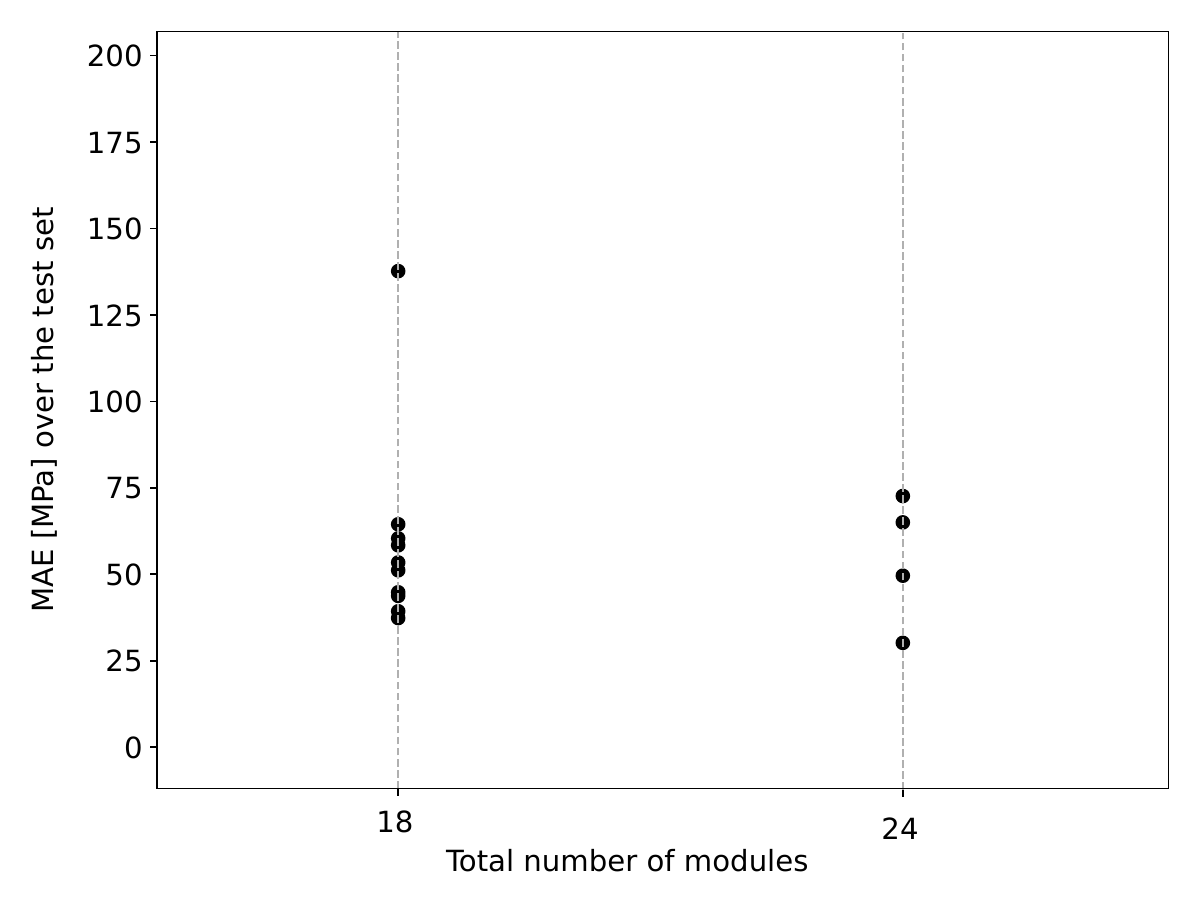}
        \label{fig:error_units_zoom}
        \caption{}
    \end{subfigure}
    \caption{Comparison of validation mean absolute errors for the mesoscale network configured with different numbers of modules in the latent space, trained across 200 training samples, and evaluated on a consistent set of 20 unseen simulations.}
    \label{fig:error_units}
\end{figure}

Using a fixed number of training samples generated from random-walk loadings, various HPRNN models are trained with different numbers of module sets. A module set combines bulk points, warp-PRNN, and weft-PRNN. Since the input strain and output stresses are six-dimensional, one module set corresponds to six units, with two units devoted to each module.
Figure \ref{fig:error_units} illustrates the effect of increasing the number of module sets from one to four. For each configuration, 10 networks with different initializations are trained and evaluated on the same test set. The results, shown in Figure \ref{fig:error_units}(b), indicate that the difference in errors between three and four module sets is minimal. Based on this observation, the configuration with three module sets (18 units in total: 3\(\times\)6 for each module) is selected as the optimal model.

\begin{figure}
    \centering
    \includegraphics[width=\linewidth]{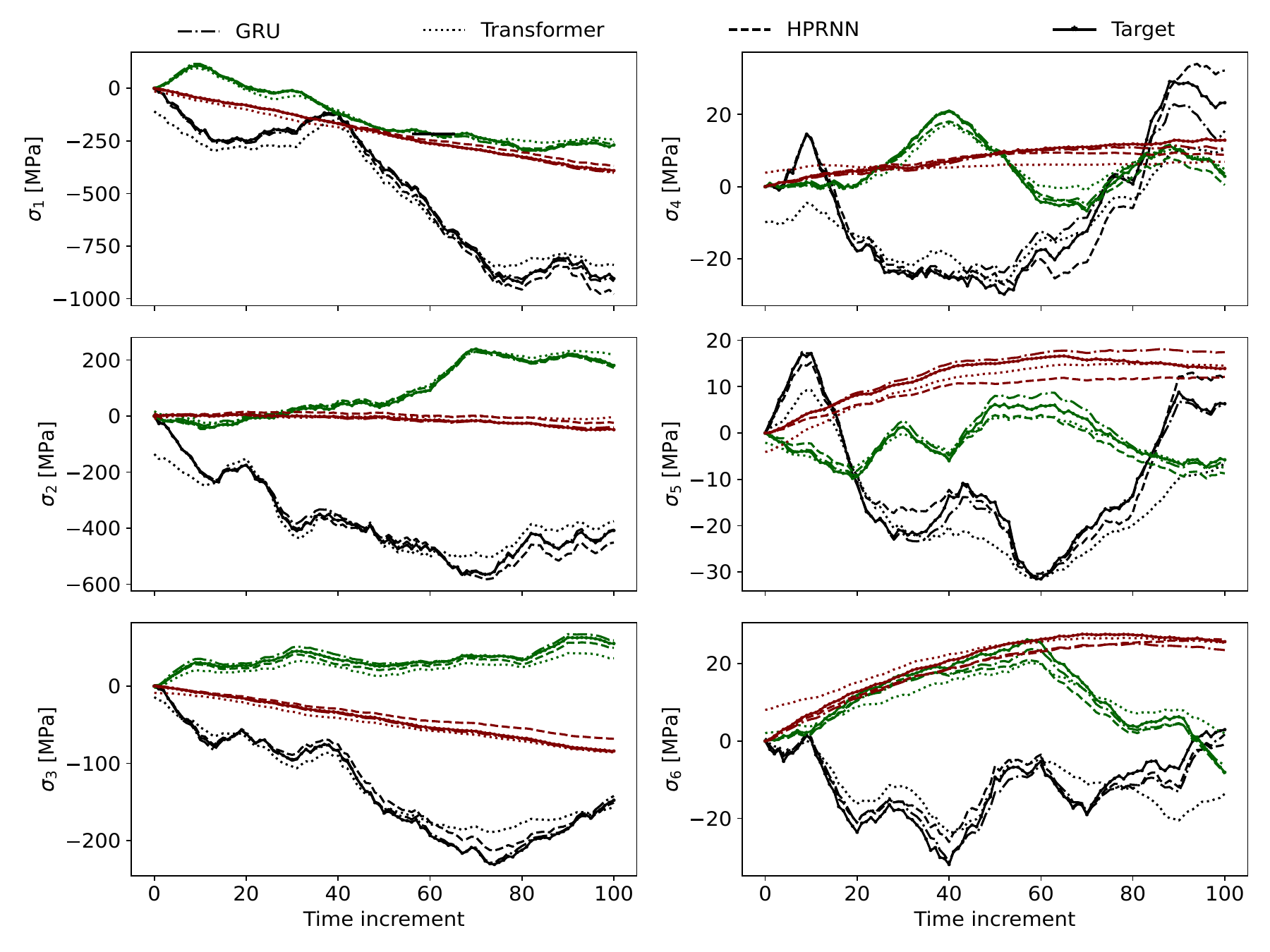} 
    \caption{Prediction of stress components by the mesoscale HPRNN, GRU, and Transformer models on three randomly selected unseen samples from random loading data, each represented by a different color. Networks predictions are plotted against simulations, with solid lines indicating the target values from high-fidelity simulations.}
    \label{fig:stresspredictionvstime}
\end{figure}

In Figure \ref{fig:stresspredictionvstime}, a randomly selected simulation from the test set of the random loading dataset is compared with the predicted values from the proposed architecture. The results demonstrate that the HPRNN model effectively captures the sequential nonlinear elasto-plastic behavior of the woven composite.

\subsection{Extrapolation capability: HPRNN vs. GRU and Transformer}
\label{extrapolation}
This section explores the extrapolation capabilities of HPRNN compared to 
GRU and transformer-based architectures. A GRU-based model, previously employed in \cite{ghane2024recurrent,ghane2024datafusion}, is used as a benchmark. The GRU model consists of three gated recurrent layers, each with a hidden size 512, and includes a dropout layer with a rate of 0.5 to reduce overfitting. The final fully connected layer maps the GRU output to the desired dimensions, processing input and output sequences of size 6. The model is trained with the AdamW optimizer (\( \text{learning rate}=0.001, \text{weight decay}=1 \times 10^{-4} \)) using a StepLR scheduler, which reduces the learning rate by \(10\%\) every five epochs. %

\begin{figure}
    \centering
    \captionsetup{justification=centering}
    \begin{subfigure}[b]{0.45\linewidth}
        \centering
        \includegraphics[width=\linewidth]{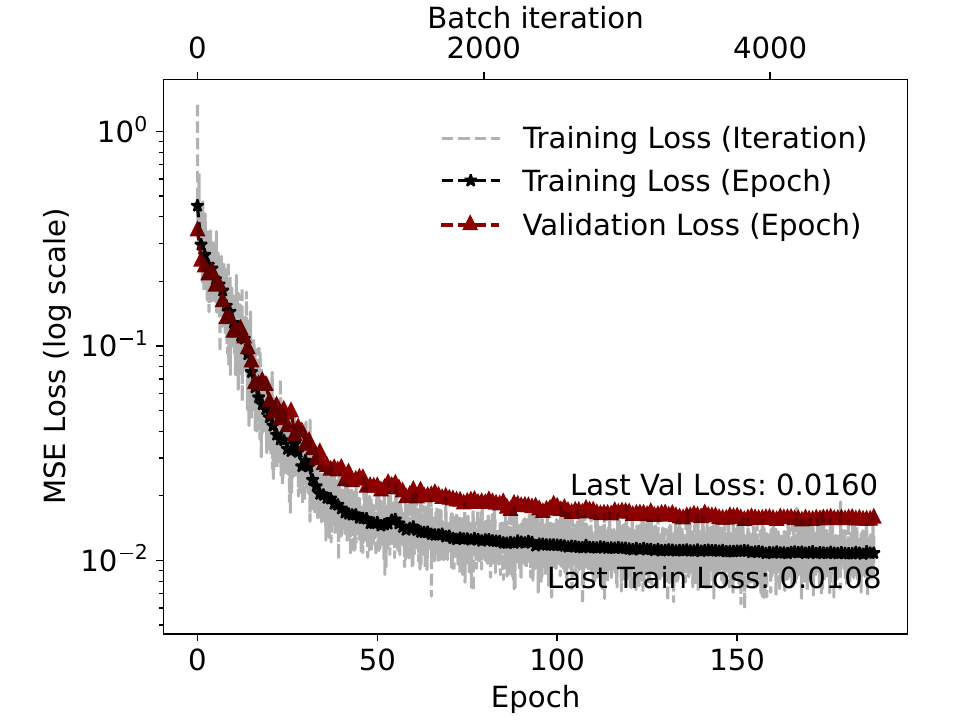}
        \caption{}
        \label{fig:gru-training}
    \end{subfigure}
    \hfill
    \begin{subfigure}[b]{0.45\linewidth}
        \centering
        \includegraphics[width=\linewidth]{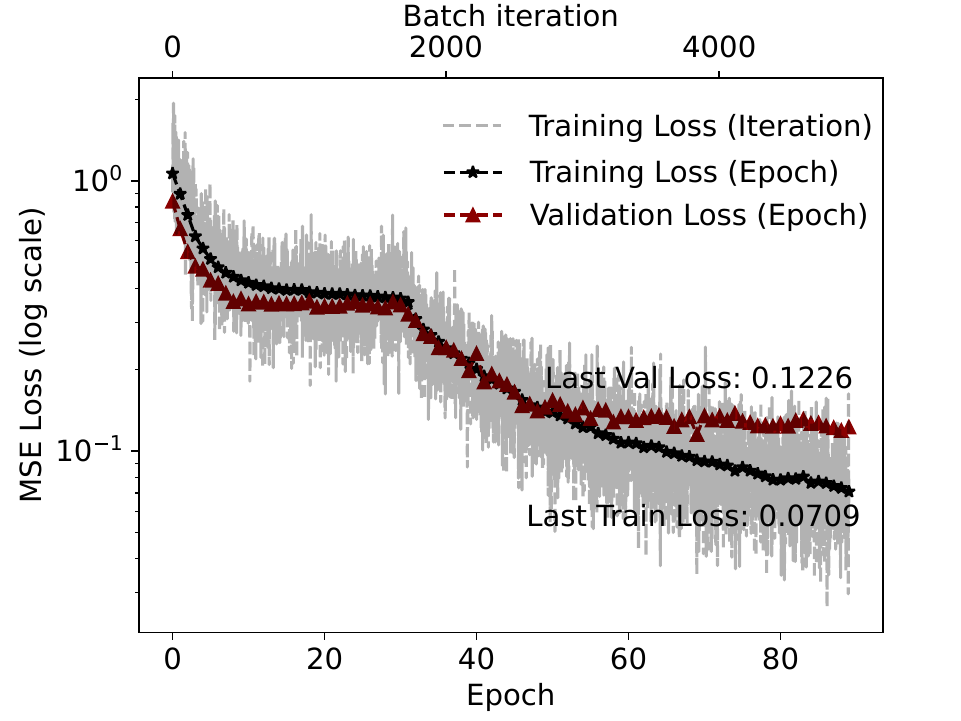}
        \caption{}
        \label{fig:transformer-training}
    \end{subfigure}
    \caption{(a) GRU and (b) transformer training curves over 200 and 437 random loading simulations, respectively. The GRU model with an optimized configuration is introduced in \cite{ghane2024datafusion}. The transformer model is adopted from the introduced configuration in \cite{zhongbo2024pre}. Both networks are trained with an early stopping criterion set to the patience of 20 epochs.}
    \label{fig:gru_Transformer-training}
\end{figure}

In comparison, a transformer-based architecture adopted from \cite{zhongbo2024pre} is implemented to evaluate its performance. The transformer model includes three encoder and decoder layers, each with a hidden size 40, and two attention heads. The feed-forward sub-layer in each block has a dimensionality of 120, ensuring adequate representational capacity for sequential data. A dropout rate of 0.1 is applied to prevent overfitting. The transformer is trained using the AdamW optimizer with a learning rate of \(0.0005\), combined with a warm-up cosine learning rate scheduler that ramps up over the first 40 epochs and decays thereafter. The MSE loss function on normalized data by RobustScaler is used to train GRU and transformer-based models as shown in Figure~\ref{fig:gru_Transformer-training}.  

\begin{figure}
    \centering
    \includegraphics[width=\linewidth]{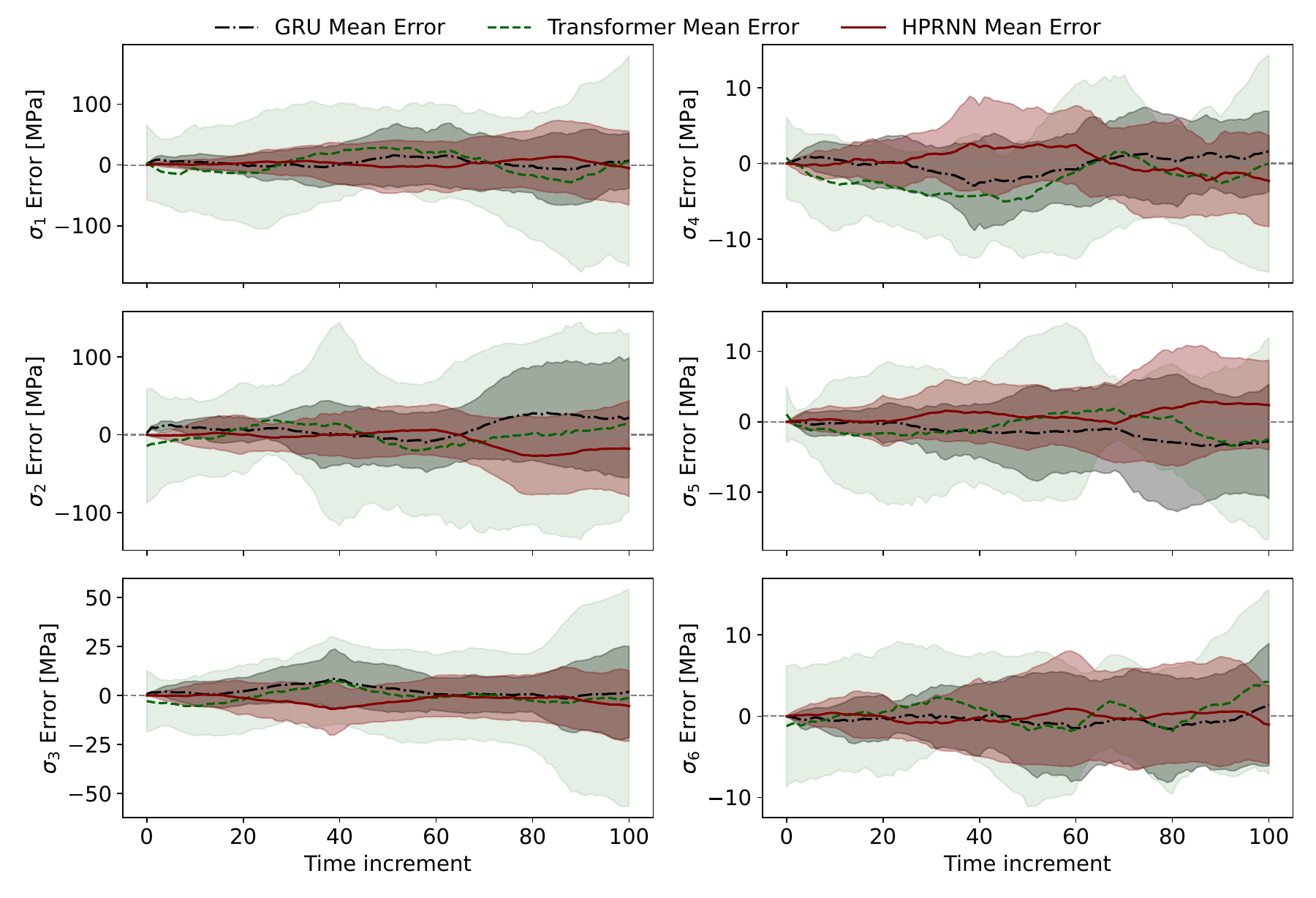}
    \caption{Mean error and error standard deviation comparison for 20 random loading samples from the test set, evaluated across three networks: GRU recurrent neural network (gray), Transformer network (green), and HPRNN (red).}
    \label{fig:error_distribution_random}
\end{figure}

Even when practically doubling the training set size from 200 to 437 samples compared to the GRU and HPRNN models, the transformer struggles to accurately predict random loading cases, as illustrated in Figure~\ref{fig:stresspredictionvstime} for three randomly selected samples from the random loading dataset. Furthermore, to evaluate the performance of the three networks across all random loading test samples, the error values relative to the target stress \((\hat\sigma_{{vM},i}^{(t)}-{\sigma}_{{vM},i}^{(t)})\) are plotted for all six stress components in Figure~\ref{fig:error_distribution_random}. The shaded regions in the figure represent errors' uncertainty (standard deviation). The transformer shows the highest error standard deviation values across time increments, even in the initial states corresponding to zero stress values, while the GRU network and HPRNN exhibit similar error distributions. However, the HPRNN performs better, particularly for out-of-plane tensile and in-plane shear stress components.

\begin{figure}
    \centering
    \includegraphics[width=\linewidth]{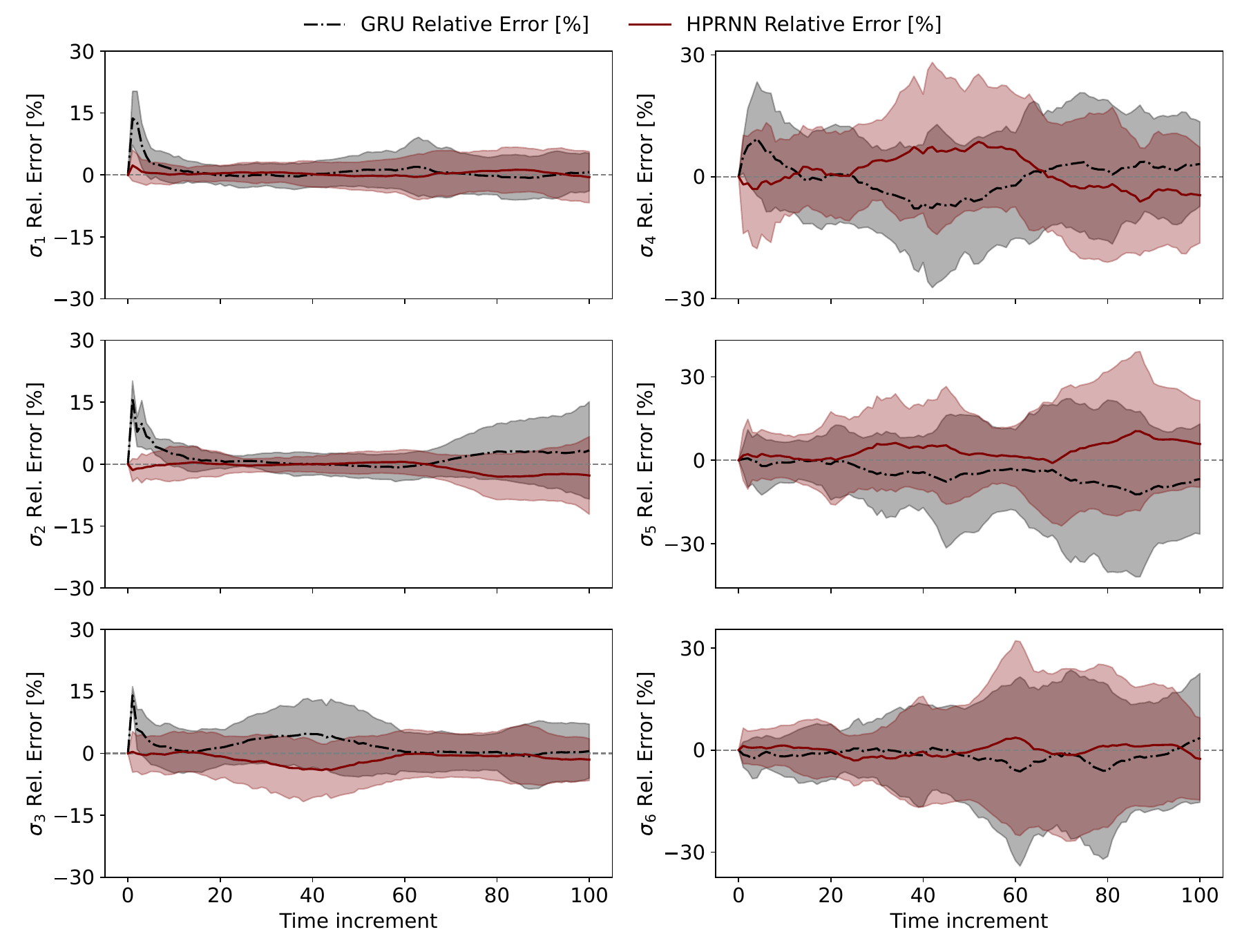}
    \caption{The relative error distribution of GRU (black) and HPRNN (red) models is normalized with respect to the absolute maximum target stress at each time increment. The shaded regions represent the standard deviation of errors over the selected test samples.}
    \label{fig:error_distribution_relative}
\end{figure}

Figure \ref{fig:error_distribution_relative} illustrates the relative error distribution of GRU and HPRNN models to facilitate a balanced comparison across different stress components, which inherently exhibit varying magnitudes. The errors are normalized with respect to the absolute maximum target stress at each time increment, enabling a more objective assessment of predictive accuracy by accounting for differences in the order of magnitude among stress components.

Both models demonstrate a strong predictive performance for normal stress components, likely due to their lower sensitivity to noise introduced during the load generation process. 
Moreover, normal stresses exhibit lower variance compared to shear stresses.

Shear stress predictions in both GRU and HPRNN models exhibit greater error variance than normal stresses. The GRU model tends to show a systematic negative bias in shear components, whereas HPRNN errors fluctuate more symmetrically. Notably, the HPRNN model demonstrates an increasing error trend in shear components over time. 

The instability in shear stress predictions suggests that caution is required when applying these models to torsional loading, or multi-axial stress scenarios. Improvements such as enhanced regularization techniques or additional feature engineering may help refine shear stress predictions and improve overall model robustness.

\begin{figure}
    \centering
    \captionsetup{justification=centering}
    \begin{subfigure}[b]{0.45\linewidth}
        \centering
        \includegraphics[width=\linewidth]{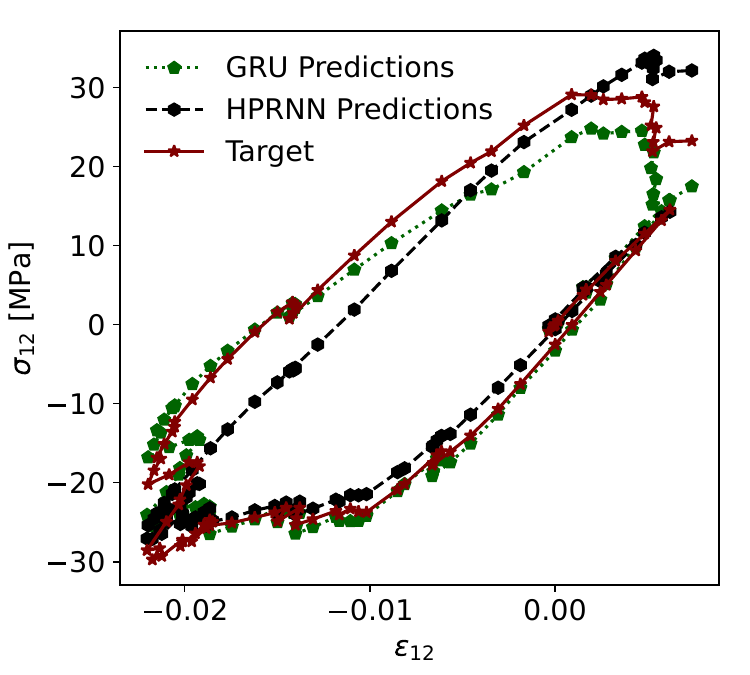}
        \caption{}
        \label{fig:stresspredictionvsstrain-a}
    \end{subfigure}
    \hfill
    \begin{subfigure}[b]{0.45\linewidth}
        \centering
        \includegraphics[width=\linewidth]{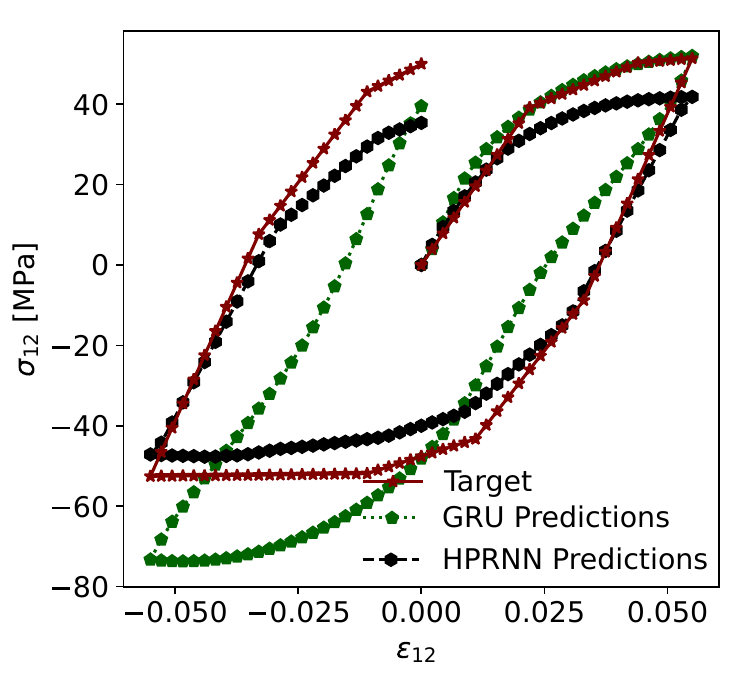}
        \caption{}
        \label{fig:stresspredictionvsstrain-b}
    \end{subfigure}
    \begin{subfigure}[b]{0.45\linewidth}
        \centering
        \includegraphics[width=\linewidth]{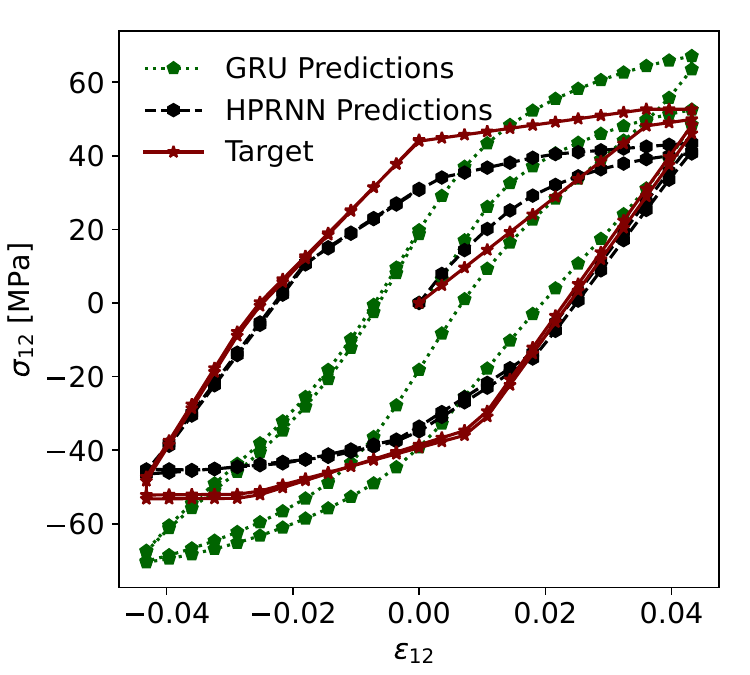}
        \caption{}
        \label{fig:stresspredictionvsstrain-c}
    \end{subfigure}
    \vspace{-0.3cm}
    \caption{In-plain stress vs. strain behavior in testing against (a) unseen random loading and extrapolation over unseen cyclic load cases. (b) one cycle, and (c) two cycles.}
    \label{fig:stresspredictionvsstrain}
\end{figure}

The GRU model achieves a performance comparable to HPRNN in predicting random load responses.
However, to further evaluate the performance of the HPRNN architecture, GRU and HPRNN are tested on cyclic loading samples described in Section~\ref{LoadGenerator} that were not part of the training set. As it was already introduced in \cite{ghane2024recurrent}, recurrent neural networks, including GRUs, struggle with extrapolation from random loads to such cyclic loading, particularly under unseen load patterns. In cyclic load cases, such as single and multiple cycles shown in Figures~\ref{fig:stresspredictionvsstrain-b} and \ref{fig:stresspredictionvsstrain-c}, the GRU model exhibits nonphysical softening behavior and inconsistent maximum and minimum stress values. This behavior persists despite normalization during training, indicating limitations in GRU’s recurrent architecture, which relies on hidden states for recurrence.

\begin{figure}
    \centering
    \includegraphics[width=0.6\linewidth]{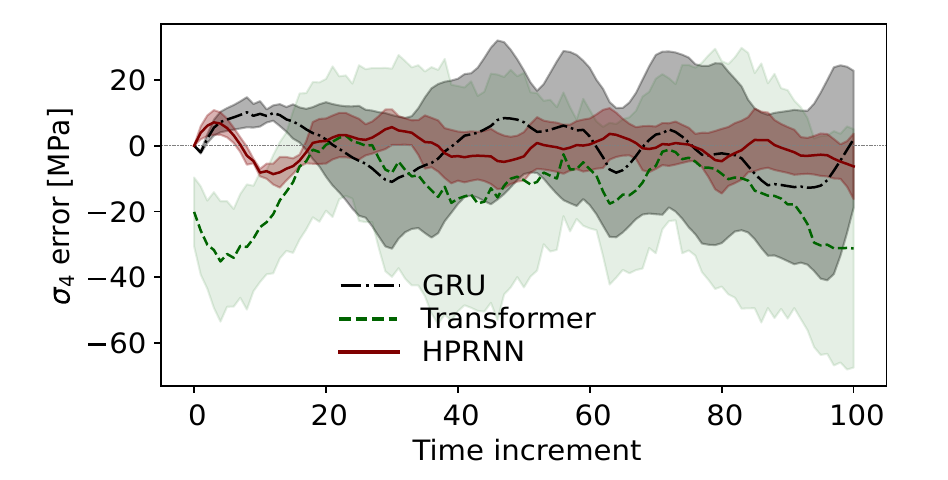}
    \caption{Mean error and error standard deviation comparison for five in-plane shear cyclic loading samples from the test set, evaluated across three networks: GRU recurrent neural network (gray), Transformer network (green), and HPRNN (red).}
    \label{fig:error_distribution_cyclic}
\end{figure}

The HPRNN, on the other hand, demonstrates superior performance in extrapolation tasks, avoiding nonphysical behaviors due to its physics-encoded recurrence mechanism. By embedding the material constitutive laws into the network, HPRNN accurately models the path-dependent elasto-plastic responses of woven composites. Figure~\ref{fig:error_distribution_cyclic} highlights its ability to maintain physically consistent predictions even under complex cyclic loading scenarios, outperforming GRU and transformer-based models. While transformers are powerful for specific sequential tasks, their inability to generalize effectively to path-dependent behaviors in this study underscores their unsuitability for woven composite modeling.

\subsection{Limitations and strengths}
\label{discussion}

Despite the advancements of the HPRNN model and its success in capturing the complex physical behavior of woven composites, the architecture faces several challenges.

HPRNN is limited to a single geometrical system (e.g., volume fraction) and requires retraining for different material configurations. In contrast, data-driven models like GRUs and LSTMs \cite{ghane2023multiscale} can generalize across various weave patterns by incorporating geometrical variables. With sufficient data, they can also capture geometric variations like fiber waviness or porosity. While HPRNN ensures physics-encoded learning, its lack of flexibility limits its applicability to diverse material systems.
Furthermore, HPRNN's complex architecture makes it's implementation less straightforward than conventional time-dependent neural networks. 

Although the HPRNN architecture improves upon GRU networks by avoiding nonphysical behavior in cyclic loading extrapolation, its predictions are not entirely accurate for all cyclic load scenarios. These discrepancies suggest a potential limitation in the current data generation strategy, specifically in using MFH for subscale transition in FFT-based data generation.

Despite these complexities, HPRNN's efficient algorithms in finite-dimensional settings, robustness in handling data scarcity, and strong generalizability offer significant potential for accelerating traditional scientific computing in applications such as computational solid mechanics.

\section{Conclusions}
\label{Conclusions}
This study presents the hierarchical physically recurrent neural network (HPRNN) as an efficient and interpretable framework for modeling woven composites. HPRNN combines data-driven learning with physical principles to predict the elasto-plastic behavior of woven composites while maintaining consistency across scales. By incorporating pretrained PRNNs for warp and weft yarns along with a matrix constitutive model, HPRNN captures detailed microscale material behavior and propagates it to the mesoscale. 
Furthermore, in contrast to conventional history-dependent neural networks, such as GRU- and transformer-based architectures, HPRNN mitigates nonphysical behaviors in extrapolation tasks by leveraging its physics-encoded recurrence design.

While transformers excel at capturing long-term dependencies in language-based data, their performance is limited when applied to path-dependent systems, such as the stress-strain behavior of the material system under study. Even with a doubled training set size, the transformer model failed to match GRU and HPRNN performance in random loading cases. Unlike RNNs or physics-encoded networks, transformers do not inherently maintain a notion of "memory" of previous states beyond what is encoded in the attention mechanism.

Key findings reveal that while the GRU network and HPRNN perform similarly under random loading conditions, their differences become evident in cyclic loading scenarios absent from the training data. HPRNN outperforms GRU by maintaining physically consistent stress-strain evolution, preventing nonphysical softening and unrealistic peak values. This advantage stems from HPRNN’s physics-encoded recurrence, which captures constitutive behavior directly from material points, unlike GRU, which relies solely on hidden states.

Future research should focus on improving HPRNN’s ability to generalize in cyclic extrapolations, potentially through transfer learning strategies, as demonstrated in prior data-driven studies \cite{ghane2023multiscale, ghane2024datafusion}. Additionally, expanding the hierarchical framework to incorporate damage mechanisms, dynamic effects, and anisotropic fiber behaviors could enhance its applicability to a broader range of materials and loading conditions. Another key direction is optimizing training efficiency through parallel processing, advanced batch handling, and streamlined computation of physics-based activation functions to enable HPRNN’s application to large-scale industrial problems.

The HPRNN framework provides a promising approach to accelerating multiscale modeling of woven composites, offering a unique advantage over conventional data-driven homogenization techniques by enabling a physics-consistent transition across micro, meso, and macro scales. This makes it a valuable tool for engineering design, material optimization, and the efficient prediction of composite behavior in real-world applications.

\section*{Acknowledgment}
\noindent Ehsan Ghane and Mohsen Mirkhalaf gratefully acknowledge financial support from the Swedish Research Council (VR grant: 2019-04715) and the University of Gothenburg. Martin Fagerström is thankful for the support through Vinnova's strategic innovation programme LIGHTer, in particular via the project LIGHTer Academy Phase 4 (grant no. 2023-01937). Iuri Roch and Marina Maia gratefully acknowledge support from the TU Delft AI Labs programme.

\bibliographystyle{unsrtnat}
\bibliography{references}  






\end{document}